\documentclass[traditabstract]{aa}

\usepackage{txfonts}
\usepackage{graphicx}
\usepackage{natbib}
\bibpunct{(}{)}{;}{a}{}{,} 
\usepackage{longtable} 

\def\solar {\ifmmode_{\mathord\odot} \else $_{\mathord\odot}$\fi}
\def\Msol {\ifmmode {\,{\it M}\solar} \else $\,M$\solar\fi}     
\def\Rsol {\ifmmode {\,{\it R}\solar} \else $\,R$\solar\fi}     
\def\Lsol {\ifmmode {\,{\it L}\solar} \else $\,L$\solar\fi}     
\def\chisq{\mbox{$\chi^2$}}
\def\Rearth{\hbox{$\mathrm{R}_{\oplus}$ }}

\newcommand{\Mearth}{M$_{Earth}$}

\begin{document}

\title{The HARPS search for southern extra-solar planets
       \thanks{Based on observations made with the HARPS instrument on the
         ESO 3.6-m telescope at La Silla Observatory under program ID
         072.C-0488, 180.C-0886, 082.C-0718, and 183.C-0437}
       }
\subtitle{XXXII. Only 4 planets in the Gl~581 system }

\author{
T. Forveille \inst{1,2}     
\and X. Bonfils \inst{1}  
\and X. Delfosse \inst{1}
\and R.~Alonso\inst{3} 
\and S. Udry \inst{3}  
\and F. Bouchy \inst{4,5}%
\and M. Gillon \inst{6}  
\and C. Lovis \inst{3}   
\and V. Neves \inst{1,7,8}  
\and M. Mayor \inst{3}      
\and F. Pepe \inst{3}       %
\and D. Queloz \inst{3}    %
\and N.C. Santos \inst{7,8}  
\and D. S\'egransan \inst{3} %
\and J.-M.~Almenara\inst{9,10,11}
\and H.J.~Deeg\inst{10,11} 
\and M.~Rabus\inst{10,11,12} 
}

\offprints{T. Forveille, \email{Thierry.Forveille@obs.ujf-grenoble.fr}}

\institute{UJF-Grenoble 1 / CNRS-INSU, 
           Institut de Plan\`etologie et d’Astrophysique de Grenoble 
               (IPAG) UMR 5274, 
           Grenoble, F-38041, 
           France
\and
           Institute for Astronomy, 
           University of Hawaii, 
           2680 Woodlawn Drive, 
           Honolulu HI 96822 USA
\and
               Observatoire de Gen\`eve, 
               Universit\'e de Gen\`eve, 
               51 ch. des Maillettes, 
               1290 Sauverny,
               Switzerland
\and  
	       Institut d'Astrophysique de Paris, 
               CNRS, Universit\'e Pierre et Marie Curie, 
               98bis Bd Arago, F-75014 Paris, France
\and
               Observatoire de Haute-Provence, 
               CNRS/OAMP, 
               F-04870 St Michel l'Observatoire, France
\and
               Universit\'e de Li\'ege, 
               All\'ee du 6 Aout 17
               Sart Tilman, Belgium
\and
               Centro de Astrof{\'\i}sica, Universidade do Porto, 
               Rua das Estrelas, 
               4150-762 Porto, Portugal
\and
               Departamento de F{\'\i}sica e Astronomia, 
               Faculdade de Ci{\^ e}ncias, Universidade do Porto,
               Rua das Estrelas, 
               4150-762 Porto, Portugal
\and
               Laboratoire d'Astrophysique de Marseille, 
               38 rue Fr\'ederic Joliot-Curie, 
               F-13388 Marseille Cedex 13, France
\and           Instituto de Astrof{\'i}sica de Canarias, 
               E-38205 La Laguna, Tenerife, Spain
\and           Dpto. de Astrof\'isica, Universidad de La Laguna, 
               38206 La Laguna, Tenerife, Spain
\and           Departamento de Astronom\'ia y Astrof\'isica, 
               Pontificia Universidad Cat\'olica de Chile, 
               Casilla 306, Santiago 22, Chile 
}

\abstract{
The Gl~581 planetary system has generated wide interest, because its 
4~planets include both the lowest mass planet known around a main 
sequence star other than the Sun and the first super-Earth planet in the 
habitable zone of its star. A recent paper announced the possible 
discovery of two additional super-Earth planets in that system, one of 
which would be in the middle of the habitable zone of Gl~581.
The statistical significance of those two discoveries has, however, 
been questioned. We have obtained 121 new radial velocity measurements
of Gl~581 with the HARPS spectrograph on the ESO 3.6~m telescope,
and analyse those together with our previous 119 measurements of that
star to examine these potential additional planets. We find that
neither is likely to exist with their proposed parameters. We 
also obtained photometric observations
with the 2.5~m Isaac Newton Telescope during a potential transit of
the inner planet, Gl~581e, which had a 5\% geometric transit 
probability. Those observations exclude transits for planet
densities under 4 times the Earth density within $-0.2$~$\sigma$ to
$+2.7$~$\sigma$ of the predicted transit center.
}

\date{}


\titlerunning{Only 4 planets in the Gl~581 system}
\authorrunning{Forveille et al.}

\maketitle

\section{Introduction}

Amongst the almost 500 planetary systems   
currently listed in the on-line Exoplanet  
Encyclopedia\footnote{http://exoplanet.eu/},
Gl~581 occupies a special place in both astronomers' and
the public's mind, as it contains the only extrasolar planet(s) of 
likely rocky composition known in the habitable zone of its star, as 
well as the lightest extrasolar planet known to this date. 
We first detected Gl~581b \citep{Bonfils2005b}, a Neptune-mass planet 
with a 5~days period, based on 20 radial velocity measurements with 
the {\it HARPS} spectrograph on the ESO 3.6~m telescope. Motivated 
in part by structure in the residuals to the \citet{Bonfils2005b} 
orbit, we then obtained a further 30 {\it HARPS} radial velocity 
measurements. Those 50 observations detected two additional planets, 
Gl~581c and Gl581~d \citep{Udry2007}, with minimum masses of 5 and 
8~\Mearth and orbital periods of 13 and 84 days. The minimum masses 
make rocky compositions most likely for both planets, and the periods
respectively locate them at the inner and outer edges of the habitable 
zone of their M3V host star. With a further 69  {\it HARPS}
measurements, for a total of 119, we finally identified a fourth 
planet \citep{Mayor2009}, Gl~581e, with a minimum mass under 2~\Mearth 
and a 3 days orbital period. To this date, Gl~581e remains the lowest 
mass planet listed in the Exoplanet Encyclopedia. Just as importantly 
for the present discussion, \citet{Mayor2009} also demonstrated that 
the 84 days \citet{Udry2007} period for Gl~581d was a one-year alias of its
true period, 67 days, marginally changing the mass of the planet
but bringing it $\sim$13\% closer to the star and further into its
habitable zone. Finally, \citet{Mayor2009} demonstrated that 
stability of Gl~581e against the gravitational influence of the 
more massive planets requires an inclination of the (assumed coplanar) 
system above 40~degrees, and therefore that the masses of the planets 
are at most 1.6 times their minimum values.

Gl~581c initially generated most excitment, since rough equilibrium 
temperature computations suggested that its surface could harbor 
liquid water, while surface water on Gl~581d would be frozen
\citep{Udry2007}. Accounting for the greenhouse effect of 
atmospheres on both planets, however,  \citet{Selsis2007} soon 
showed that an extremely high albedo would be needed for the 
surface of Gl~581c to escape water boilout, while realistically 
high concentrations of atmospheric CO$_2$ might be sufficient 
to keep Gl~581d from freezing out.
The latter planet is therefore by far the better candidate for 
habitability, a conclusion only made stronger by the $\sim$25\% 
higher stellar flux at its closer \citet{Mayor2009} orbital distance
\citep[e.g.][]{Wordsworth2011}.

Most recently, \citet{Vogt2010} announced the discovery of two
additional planets in the Gl~581 system, based on a joint analysis
of the 119 \citet{Mayor2009} {\it HARPS} radial velocities
and 122 previously unpublished {\it Keck HIRES} measurements.
The two additional planets, Gl~581f and Gl581g, would have 
masses of 7 and 3 Earth masses, and periods of 433 and 33 days
That announcement raised considerable interest, since Gl~581g
would, if confirmed, sit in the middle of the habitable zone of 
the system, when Gl~581d instead is close to the cold edge of that
zone. While Gl~581d's habitability depends on a thick enough atmosphere 
producing both a strong greenhouse effect and a vigorous dayside 
to nightside circulation, Gl~581g would likely be
habitable for a broader range of atmospheric parameters. That planet 
has therefore already been the focus of significant theoretical 
habitability work, with uniformly positive conclusions 
\citep[e.g.][]{Pierrehumbert2011,vonBloh2011,Heng2011}.
The \citet{Vogt2010} 5$^{th}$ and 6$^{th}$ planets however 
are detected at relatively modest significance level. Both
detections have therefore been questioned on purely statistical 
grounds, first by \citet{Andrae2010}, and more recently by 
\citet{Tuomi2011} and \citet{Gregory2011}. \citet{Anglada2011},
on the other hand, conclude with some tentative support 
for Gl~581f, though none for Gl~581g. Beyond purely statistical 
considerations, merging datasets from independent instruments 
or analyses necessarily exposes to some risk of being misled by 
subtle low-level systematics in one or the other, and more so when 
none of the individual datasets detects the signal of interest. 
The two additional planets, and particularly the potentially habitable 
Gl~581g, have therefore generally been received with as much 
caution as interest. 

As we announced  at a conference soon after the \citet{Vogt2010} 
manuscript first became public \citep{Pepe2011}, we had continued 
observing Gl~581 after the publication of \citet{Mayor2009}. Those
continuing observations were motivated largely by the dynamical 
stability island in the middle of the habitable zone where 
\citet{Vogt2010} announced their discovery of Gl~581g, together
with speculations that planetary systems might be ``dynamically full''.
At that time, we had obtained $\sim$60 additional HARPS measurements 
and announced that we saw no evidence for either f or g. We have 
since then observed the Gl~581 system with HARPS for one 
additional year, and have now obtained a total of 121 radial 
velocity measurements above 
the 119 discussed in \citet{Mayor2009}, again approximately 
doubling the number of HARPS measurements. 
We present the HARPS measurements dataset in 
Section~2, and discuss in Section~3 the limits which it sets 
on planets beyond the 4 discussed in \citet{Mayor2009}. 
Additionally, we obtained ground-based photometry, discussed in 
Section~4, to search for transits of the 3-days Gl~581e.

\section{HARPS observations} 

\begin{table}
\centering
\caption{
\label{Table_stellar}
Observed and inferred stellar parameters for Gl~581}
\begin{tabular}{l@{}lcl}
\hline
 \multicolumn{2}{l}{\bf Parameter}
& \multicolumn{1}{c}{\bf Gl~581} 
& {\bf Reference} \\
\hline
Spectral Type &          & M3V               & \citet{Hawley1996} \\
V             &          & $10.55 \pm 0.01$  & \citet{Mermilliod1997} \\
B$-$V         &          & $1.60 \pm 0.01$   & \citet{Mermilliod1997} \\
$\pi$         &[mas]     & $160.91 \pm	2.62$ & \citet{VanLeeuwen2007}\\
Distance      &[pc]      & $ 6.21 \pm 0.10$ & \citet{VanLeeuwen2007}\\
$M_V$         &          & $11.58 \pm 0.03$ & \\
K             &          & $5.85 \pm 0.03$ & \citet{Leggett1992}\\
$M_K$         &          & $6.88 \pm 0.04$ & \\
$L_\star$      & [$\mathrm{L_\odot}$] &  $0.013$ & \citet{Delfosse1998}\\
$L_x/L_{bol}$  &          &  $<5.10^{-6}$ & \citet{Delfosse1998}\\
$[Fe/H]$      &               & $ -0.22 $ & \citet{Schlaufman2010} \\
$M_\star$      & [$\Msol$]    & $ 0.31 \pm 0.02$ & \citet{Bonfils2005}\\
$R_\star$      & [$\Rsol$]    & $0.29$ & \citet{Bonfils2005} \\
$v\sin i$     & [km\,s$^{-1}$] & $ < 1 $ & \citet{Udry2007} \\
P$_{rot}$      & days         & $94.2 \pm 1.0$  & \citet{Vogt2010}  \\
v$_{rot}$      & [km\,s$^{-1}$]& 0.16  & This paper  \\
age & [Gyr] & $>2$ & \citet{Bonfils2005}  \\
\hline
\end{tabular}
\end{table}

Table~\ref{Table_stellar} summarizes the stellar parameters of
Gl~581, which we discussed at some length in our previous 
publications \citep{Bonfils2005,Udry2007,Mayor2009}. The most 
important update over those is that \cite{Vogt2010} used differential 
photometry to identify the likely rotation period of the
star, 94 days. From that period and the 0.29~$\Rsol$ radius
of Gl~581, one infers an equatorial rotation velocity of
just 0.16~km\,s$^{-1}$, well under the 1~km\,s$^{-1}$ spectroscopic 
upper limit on $v\sin i$ . This reinforces previous conclusions
that Gl~581 is a very slow rotator, and is consistent to its
belonging to the quietest magnetic activity quartile of the
HARPS M~dwarf sample.

\begin{figure}
\centering
\includegraphics[width=0.9\linewidth]{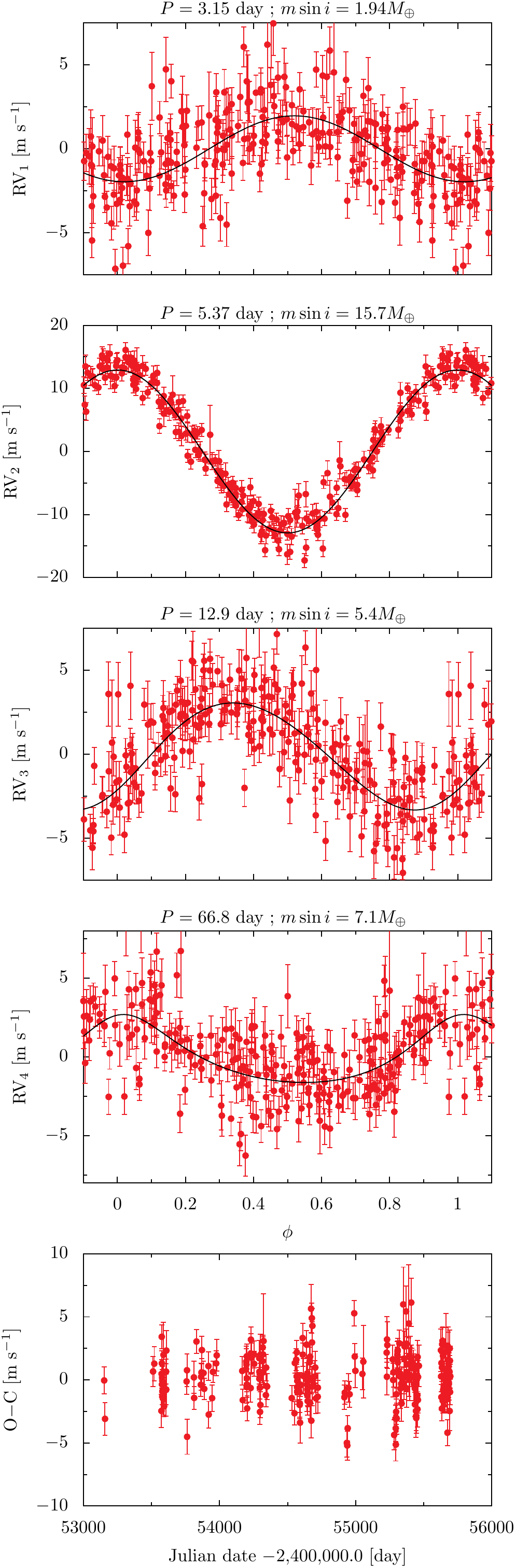}
       \caption{
        HARPS radial velocities of Gl~581 phased to the periods
        of each of Gl~581b to Gl~581e, with the best fits for the 
        other 3 planets subtracted, and residuals to the 4-planets
        fit as a function of time. The planets are ordered by increasing
        period (Gl~581e (top), Gl 581b, Gl~581c, Gl~581d,
        and residuals (bottom).
        The adjusted Keplerian orbit is overlaid.
       }
       \label{Fig_Orbit}
\end{figure}

Our observing procedure for Gl~581 is, similarly, presented in 
some detail in our previous papers, and only summarized here. 
Briefly, we obtain 15~mn exposures of 
Gl~581 with HARPS \citep[High Accuracy Radial velocity Planet 
Searcher][]{Mayor2003}, and choose to keep the calibration fiber
of the spectrograph dark. While simultaneous exposures of a 
Thorium-Argon lamp spectrum on that fiber and a stellar spectrum 
on the object fiber provide best velocity stability for HARPS
\citep[e.g.][]{Lovis2006,Mayor2009}, 
our 15~mn exposures of the V~=~10.5 Gl~581 do not reach the sub-m/s 
stability floor of the spectrograph. We can therefore rely on the excellent 
instrumental stability (nightly instrumental drifts $<$~1\,m~s$^{-1}$) 
of HARPS and obtain separate calibration exposures. This produces 
cleaner stellar spectra, suitable for quantitative spectroscopic 
analyses. We obtained 121 new HARPS measurements, for a total of 
240 with the \citep{Mayor2009} included, with a median S/N ratio 
of 46 per pixel at 550~nm. 
After accounting for observational 
overheads and for the average weather statistics at La Silla 
observatory, our observational investment on this one system
now amounts to approximately 10 nights. 

The radial velocities (Table~\ref{TableRV}, only available
electronically) were obtained with the standard HARPS data 
reduction pipeline, based on the cross-correlation with a 
stellar template and the precise nightly wavelength 
calibration with ThAr spectra \citep{Lovis2007}. For homogeneity, 
and to benefit from the continuous improvements 
to the HARPS pipeline, all measurements have been reprocessed with 
its latest version. The velocities listed in Table~\ref{TableRV} for 
the 119 measurements in common with \citet{Mayor2009} therefore 
differ slightly from the values listed in that paper, 
though usually by a small fraction of their stated standard errors.
The velocities have a median internal error of only 1.15~m\,s$^{-1}$, 
which includes the 0.3~m\,s$^{-1}$ nightly zero-point calibration 
uncertainty \citep{Lovis2007}, the  0.3~m\,s$^{-1}$ rms calibration 
drift during a night, and the photon noise computed from the full 
Doppler information content of the spectra \citep{Bouchy2001}. The 
latter dominates the error budget for this moderately faint source. 

\section{Orbital analysis}
\begin{figure}
\centering
\includegraphics[width=0.9\linewidth]{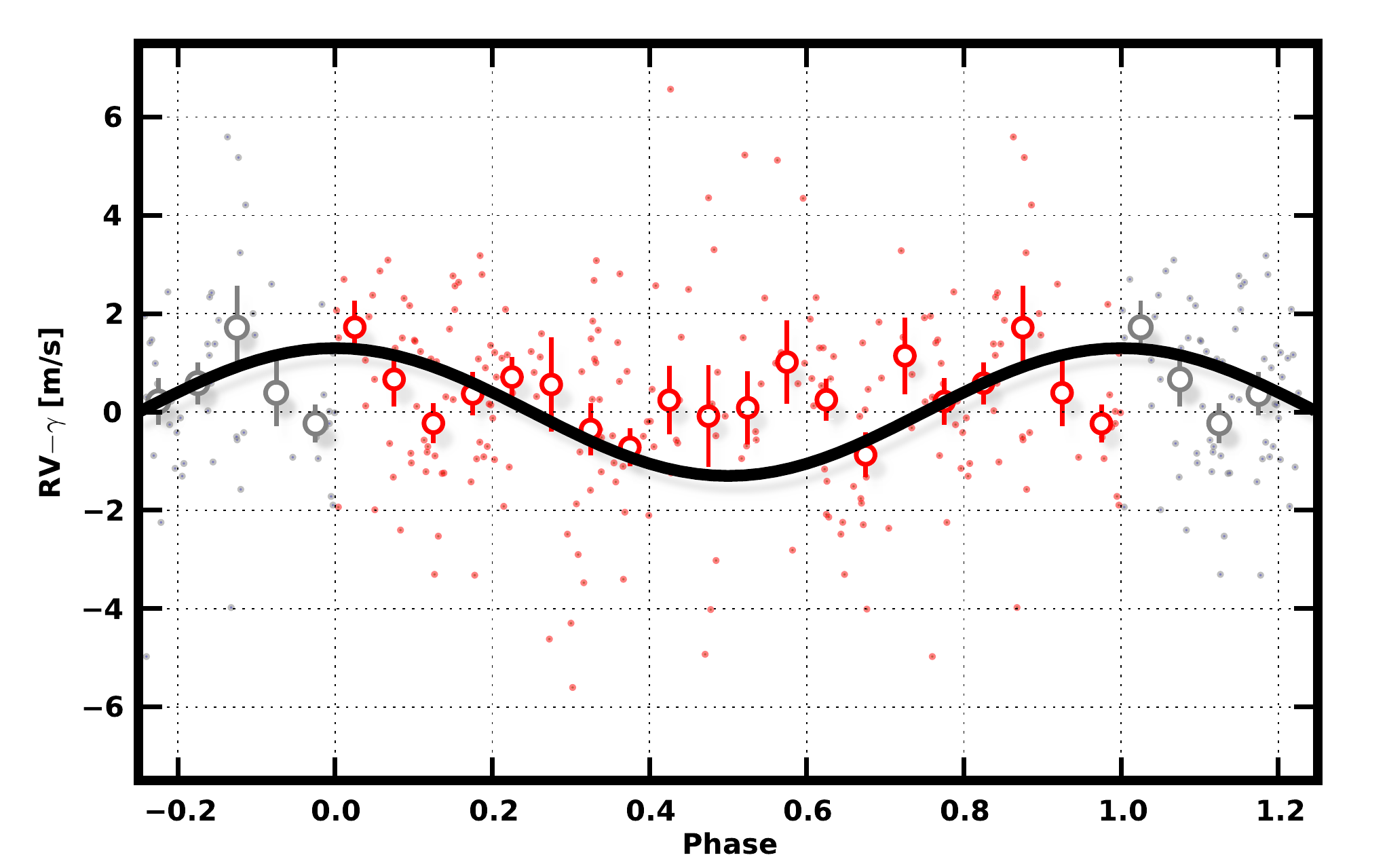}
\includegraphics[width=0.9\linewidth]{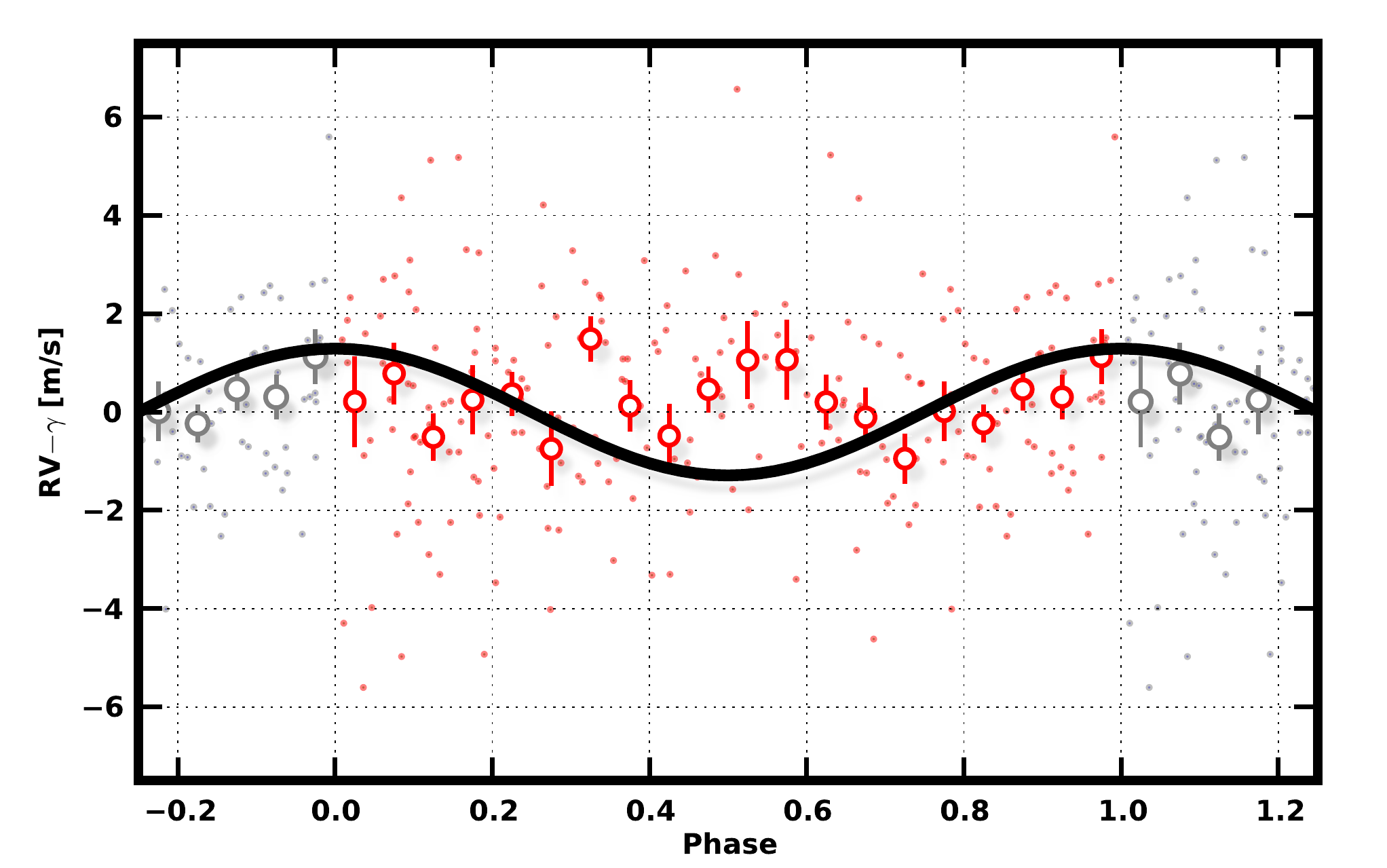}
       \caption{
        Residuals of the HARPS radial velocity measurements of 
        Gl~581 to the 4-planets Keplerian model 
        (Table~\ref{Table_Orbit}), phased to the 433 days period
        of Gl~581f (top) and and to the 36.6~days periods of
        Gl~581g (bottom). The dots represent the individual residuals, 
        and the open circles their medians over 0.05 phase bins, 
        with vertical errorbars showing the
        standard error of their mean. The thick curves represent
        the \citet{Vogt2010} orbits for both planets.
        }
       \label{Fig_Gl581fg}
\end{figure}

Based on the nominal standard errors of the two datasets,
1.70~m/s for the \citet{Vogt2010} HIRES velocities and 1.15~m/s 
for our HARPS measurements, the 121 new HARPS measurements have 
twice the statistical weight of the 122 HIRES velocities which 
\citet{Vogt2010} analysed together with the 119  \citet{Mayor2009} 
measurements. Additionally, \citet{Gregory2011} find evidence for 
$\sim$1.8 m/s of additional Gaussian noise ('jitter') above the 
nominal error bars of the \citet{Vogt2010} HIRES velocities 
and none for the \citet{Mayor2009}. This would reduce the
relative information content of the HIRES measurements by another 
factor of two. Based on their higher statistical weight, and
to retain the benefits of a homogeneous dataset, we limit our 
analysis to the HARPS measurements. One price to pay for this 
homogeneity is a shorter time span, 7~years instead of 
12 years with the HIRES data included. The HARPS measurements however
have good phase coverage for both 33 and 433~days periods 
(Fig.~\ref{Fig_Gl581fg}), and therefore provide good diagnostic 
power on their own for Gl~581f and g. 

\begin{table*}[t!]
\caption{\label{Table_Orbit} Keplerian (top) and circular (bottom) orbital 
  models of the Gl\,581 planetary system.}
\begin{tabular}{l l l c c c c}
  \multicolumn{7}{c}{\bf\large 4 Keplerian orbits}  \\
  \hline\hline
  \multicolumn{2}{l}{\bf Parameter} &\hspace*{2mm} 
  & \bf Gl\,581\,e  & \bf Gl\,581\,b  & \bf Gl\,581\,c & \bf Gl\,581\,d \\
  \hline
  $P$ & [days] & & 3.14945 $\pm$ 0.00017 & 5.36865 $\pm$ 0.00009 & 12.9182 $\pm$ 0.0022 & 66.64 $\pm$ 0.08\\
  $T0$ & [JD-2400000] & & 54750.31 $\pm$ 0.13 & 54753.95 $\pm$ 0.39 & 54763.0 $\pm$ 1.6 & 54805.7 $\pm$ 3.4 \\
  $e$ & & & 0.32 $\pm$ 0.09 & 0.031 $\pm$ 0.014 & 0.07 $\pm$   0.06 & 0.25 $\pm$   0.09  \\
  $\omega$ & [deg] & & 236 $\pm$ 17  & 251 $\pm$ 26 & 235 $\pm$ 44 & 356 $\pm$  19\\
  $K$ & [m s$^{-1}$] & & 1.96 $\pm$ 0.20 & 12.65 $\pm$ 0.18 & 3.18 $\pm$ 0.18 & 2.16 $\pm$ 0.22\\
  $V$ & [km s$^{-1}$] & & \multicolumn{4}{c}{-9.2060 $\pm$ 0.0004}  \\
  $f(m)$ & [10$^{-12} M_{\odot}$] & & 0.21 & 112.46 & 4.28 & 6.28\\
  $m \sin{i}$ & [$M_{\oplus}$] & & 1.95 & 15.86 & 5.34 & 6.06 \\
  $a$ & [AU] & & 0.028 & 0.041 & 0.073 & 0.22 \\
  \hline
  $N_{\mathrm{meas}}$ & & & \multicolumn{4}{c}{240} \\
  {\it Span} & [days] & & \multicolumn{4}{c}{2543} \\
  $\sigma$ (O-C) & [ms$^{-1}$] & & \multicolumn{4}{c}{1.79} \\
  $\chi^2_{\rm red}$ & & & \multicolumn{4}{c}{2.57} \\
\end{tabular}
\begin{tabular}{l l l c c c c}
  \hline
  \multicolumn{7}{c}{\bf\large 4 circular orbits}  \\
  \hline\hline
  \multicolumn{2}{l}{\bf Parameter} &\hspace*{2mm} 
  & \bf Gl\,581\,e  & \bf Gl\,581\,b  & \bf Gl\,581\,c & \bf Gl\,581\,d \\
  \hline
  $P$ & [days] & & 3.14941 $\pm$ 0.00022 & 5.36864 $\pm$ 0.00009 & 12.9171 $\pm$ 0.0022 & 66.59 $\pm$ 0.10\\
  $T$ & [JD-2400000] & & 54748.243 $\pm$ 0.056 & 54750.199 $\pm$ 0.012 & 54761.03 $\pm$ 0.11 & 54806.8 $\pm$ 1.0  \\
  $K$ & [m s$^{-1}$] & & 1.754 $\pm$ 0.180 & 12.72 $\pm$ 0.18 & 3.21 $\pm$ 0.18 & 1.81 $\pm$ 0.19\\
  $V$ & [km s$^{-1}$] & & \multicolumn{4}{c}{-9.2060 $\pm$ 0.0001}  \\
  $f(m)$ & [10$^{-12} M_{\odot}$] & & 0.18 & 114.54 & 4.46 & 4.11\\
  $m \sin{i}$ & [$M_{\oplus}$] & & 1.84 & 15.96 & 5.41 & 5.26 \\
  $a$ & [AU] & & 0.028 & 0.041 & 0.073 & 0.22 \\
  \hline
  $N_{\mathrm{meas}}$ & & & \multicolumn{4}{c}{240} \\
  {\it Span} & [days] & & \multicolumn{4}{c}{2543} \\
  $\sigma$ (O-C) & [ms$^{-1}$] & & \multicolumn{4}{c}{1.86} \\
  $\chi^2_{\rm red}$ & & & \multicolumn{4}{c}{2.70} \\
  \hline

\end{tabular}
\end{table*}
 
Since all analyses of the Gl~581 system agree on the broad 
characteristics of planets b to e, we start by adjusting
a 4-planets Keplerian orbital model to the HARPS measurements. 
We then mostly work from its residuals. The resulting orbital 
model (Table~\ref{Table_Orbit}, top, and Figure~\ref{Fig_Orbit}) is 
generally consistent with previous determinations, but has 
improved errorbars and provides 
better constrained ephemerides at recent and future dates. 
The rms amplitude of the residuals from the orbital model is 
1.8~m\,s$^{-1}$ significantly above the 1.15~m\,s$^{-1}$ average 
photon noise. The square root ot the reduced ${\chi}^2$ of 
the fit is consequently 1.6, and the radial velocities thus
contain either information beyond the detected planet or excess
noise, instrumental or astrophysical. The largest residuals, 
such as those which stand out at phases 0.5 to 0.6 in the 
Gl~581b panel of Figure~\ref{Fig_Orbit}, correspond to 
spectra with low S/N ratio (under 35, compared to a median 
of 46), obtained through either clouds or degraded seeing.
Ignoring those measurements produces visually more pleasing
figures, but leaves the orbital parameters essentially
unchanged and only modestly lowers the reduced ${\chi}^2$
of the least square fit. We chose to retain them, for the
sake of simplicity.

The adjusted eccentricities are small for all 4 planets, with a
formally highest significance of 3.5$\sigma$ for the 3~days
Gl~581e, which is subject to the strongest tidal forces and
least expected to have high eccentricity.
We therefore experimented with adjusting circular orbits
for all 4 planets. The resulting orbital model
(Table~\ref{Table_Orbit}, bottom) has a slightly higher 
reduced ${\chi}^2$, 2.70 compared to 2.57 for the Keplerian 
model. We adopt the latter as our baseline, but repeated
the analyses described below using the residuals of
circular orbits, with identical conclusions. The choice 
is thus of no consequence for the following discussion.

Fig.~\ref{Fig_Gl581fg} shows the residuals of the 4-planets Keplerian
model phased to the periods of the \citet{Vogt2010} Gl~581f and 
Gl~581g, as well as the medians of those residuals over 0.05 phase
bins. The binned residuals shows no evidence for the $\pm$1.3~m/s
signal expected from the two planets and overlayed as thick curves.
To quantify that statement, we adjust a 6-planet model with the 
the orbital elements of two planets fixed to the \citet{Vogt2010}
values for Gl~581f and Gl~581g. That model has a reduced ${\chi}^2$
of 4.54, well above the 2.57 of the 4-planets model, or equivalently
has 75\% higher residuals. A 6-planets
model with only the periods of Gl~581f and Gl~581g fixed to their
\citet{Vogt2010} values does produce a similar reduced ${\chi}^2$
to the 4-planets model, but it has amplitudes under 40~cm/s for
both planets and phases which do not match \citet{Vogt2010} either.
We can therefore safely conclude that the HARPS data excludes
Gl~581f and Gl~581g existing with their \citet{Vogt2010} orbital 
elements.

\begin{figure}
\centering
\includegraphics[width=0.9\linewidth]{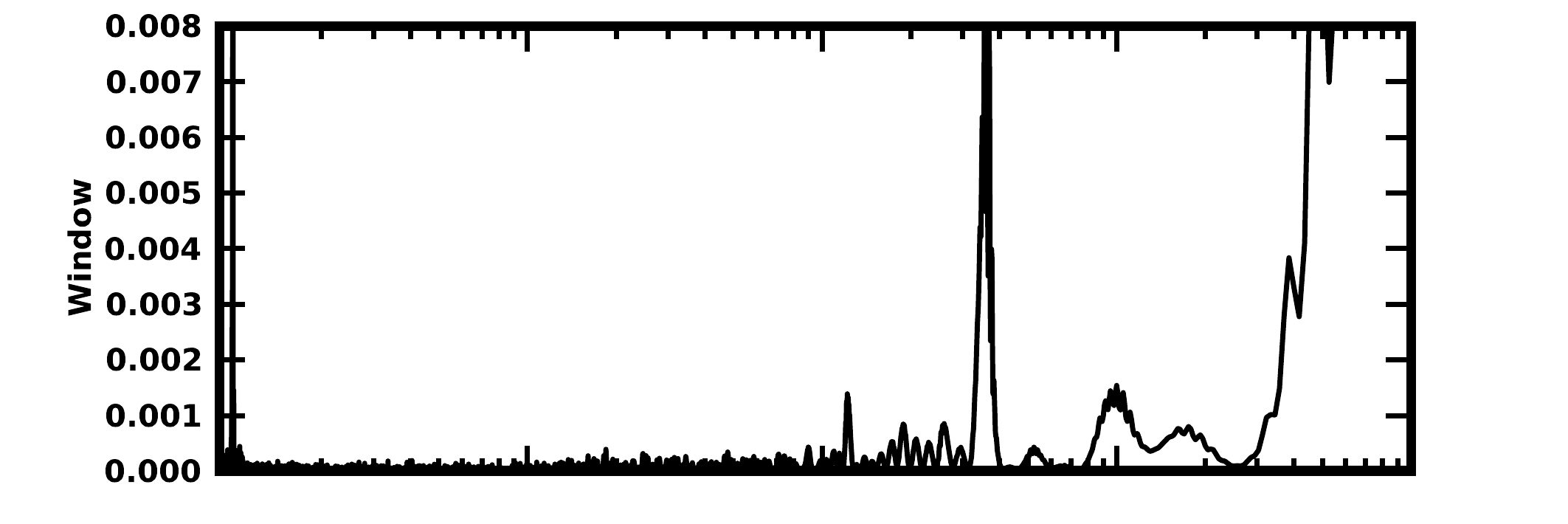}
\includegraphics[width=0.9\linewidth]{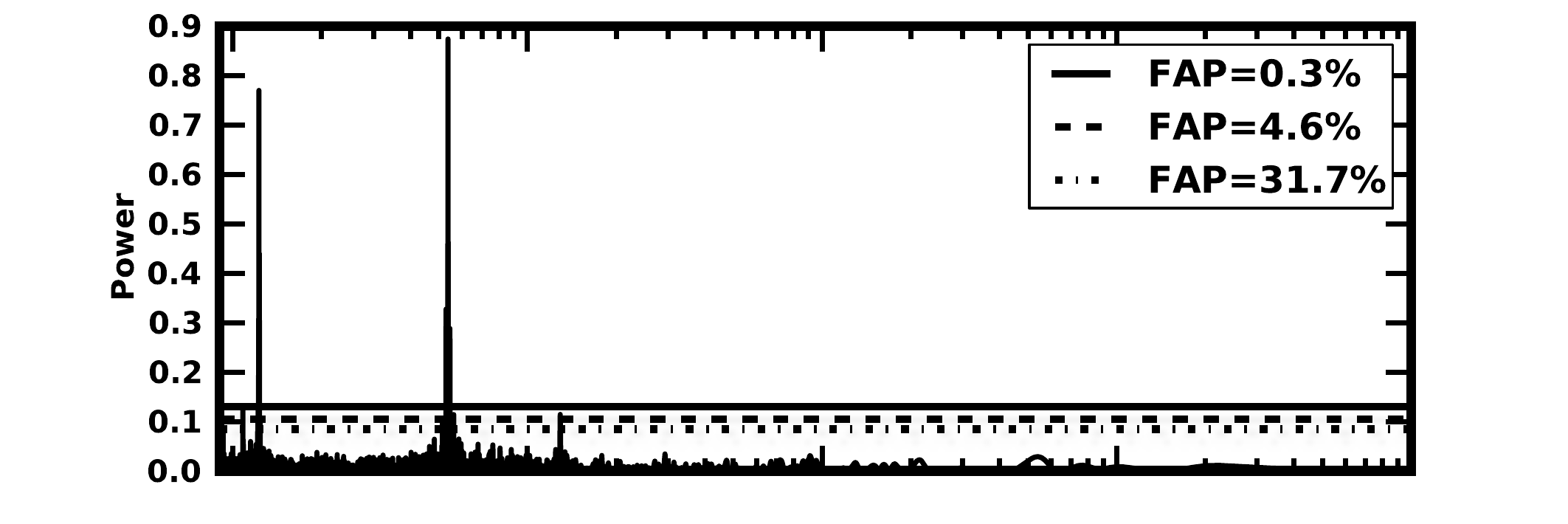}
\includegraphics[width=0.9\linewidth]{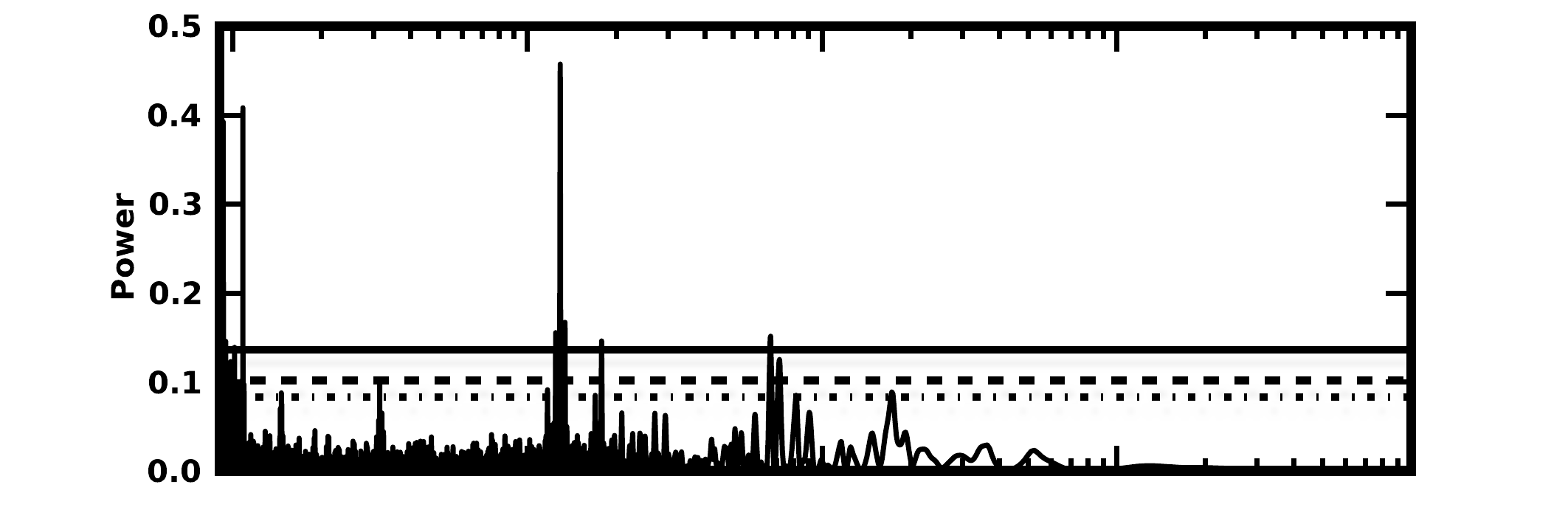}
\includegraphics[width=0.9\linewidth]{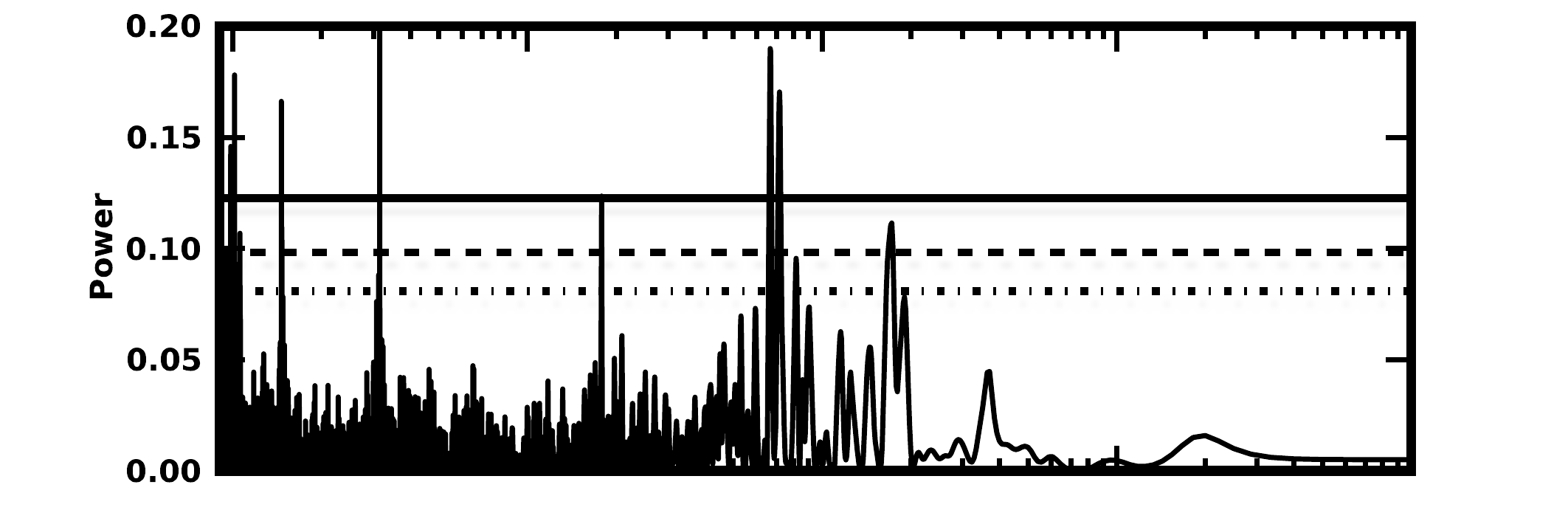}
\includegraphics[width=0.9\linewidth]{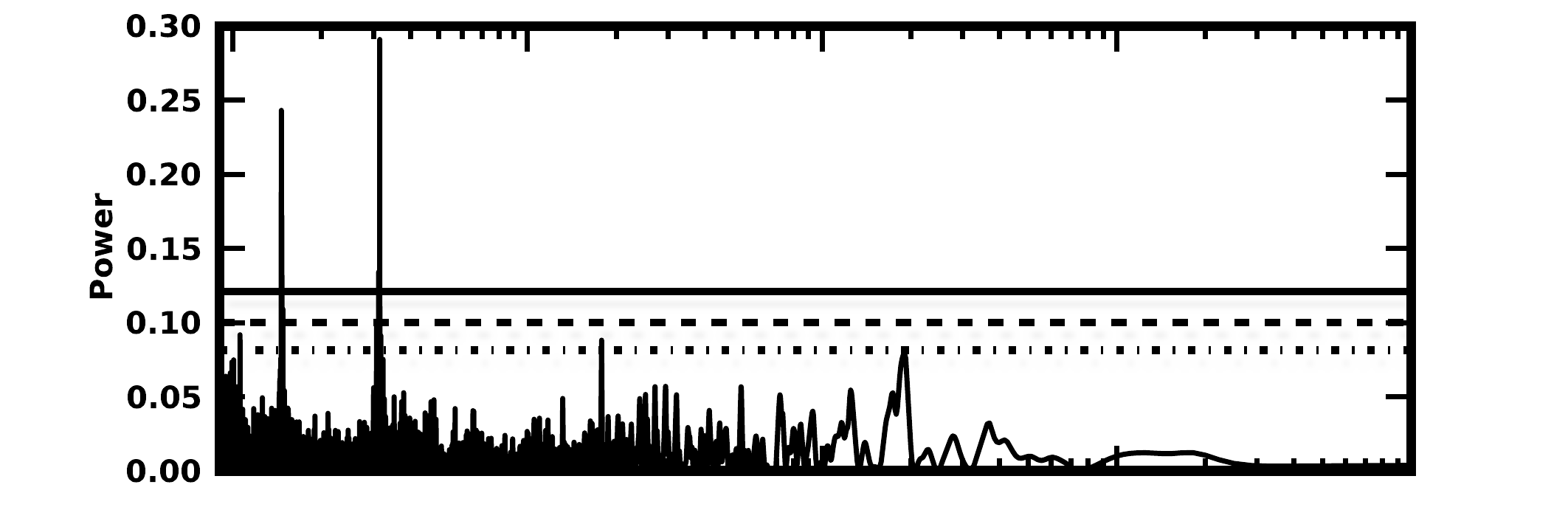}
\includegraphics[width=0.9\linewidth]{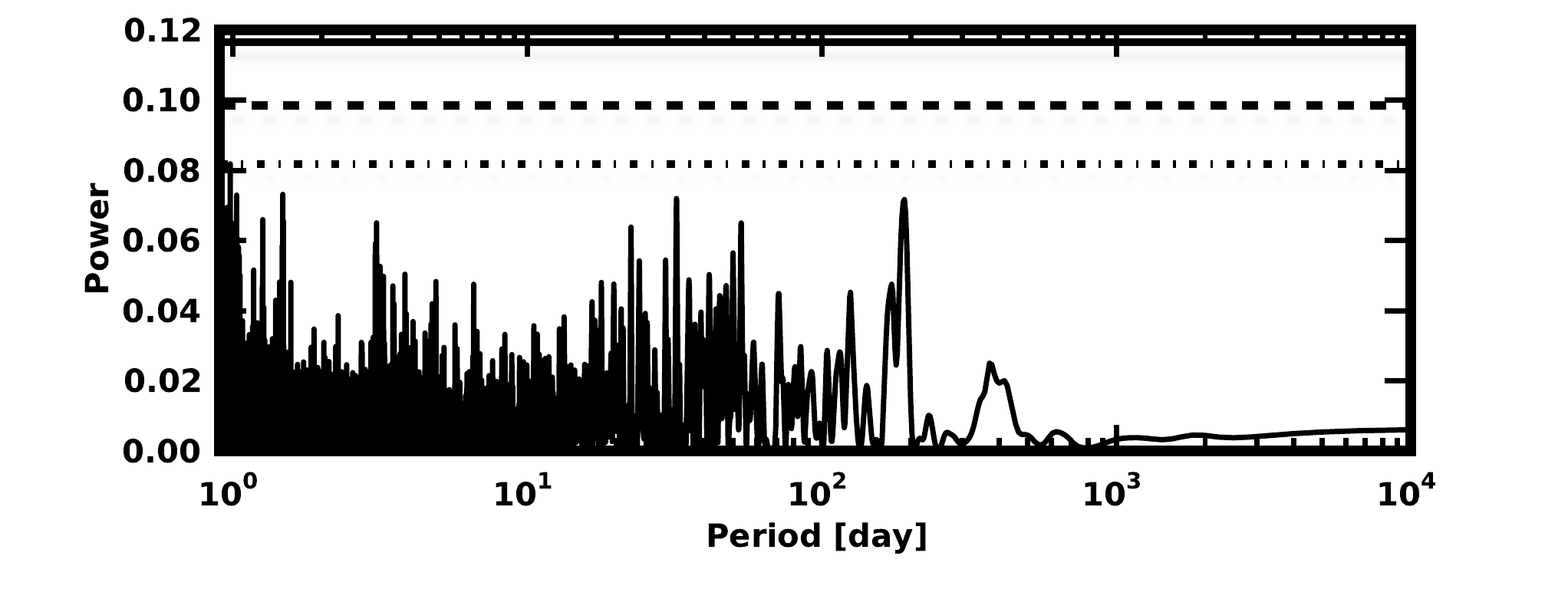}
\caption{
  Periodograms of the HARPS radial velocity
  measurements of Gl~581, with horizontal lines marking the False Alarm
  Probabilities corresponding to 1, 2 and 3~$\sigma$ thresholds for a 
  Gaussian statistics. {\it Panel~1 (top)}: Window function of the
  measurements 
  {\it Panel~2}: Periodogram of the HARPS
  velocities. The peaks corresponding to Gl~581b (P=5.36d) and its alias
  with one sidereal day are prominent, and that for  Gl~581c 
  (P=12.9d) is easily seen. {\it Panel 3}: periodogram of the 
  HARPS velocities corrected from the variation due to Gl~581b. The
  peak corresponding to Gl~581c (P=12.9d) becomes prominent, and the
  peaks corresponding to Gl~581d (P=67d) and Gl~581e (P=3.15 days)
  and their aliases can be distinguished. {\it Panel 4}:
  Removing the variability due to planets b and c makes the 
  peaks due to the P=3.15d Gl~581e (and its alias with the
  a one sideral day period) and to the P-67~days Gl~581d (and its
  one year alias). 
  {\it Panel  5} : Periodogram after removing the effect of planets b, c and
  d. The peaks due to Gl~581e (and its one-day aliases) are evident. 
  {\it Panel 6 (bottom)}: Periodogram of the residuals of the 4-planets
  orbital model, with no additional significant peak.  }
  \label{Fig_Periodogram}
\end{figure}
\begin{figure}
\centering
\includegraphics[width=0.9\linewidth]{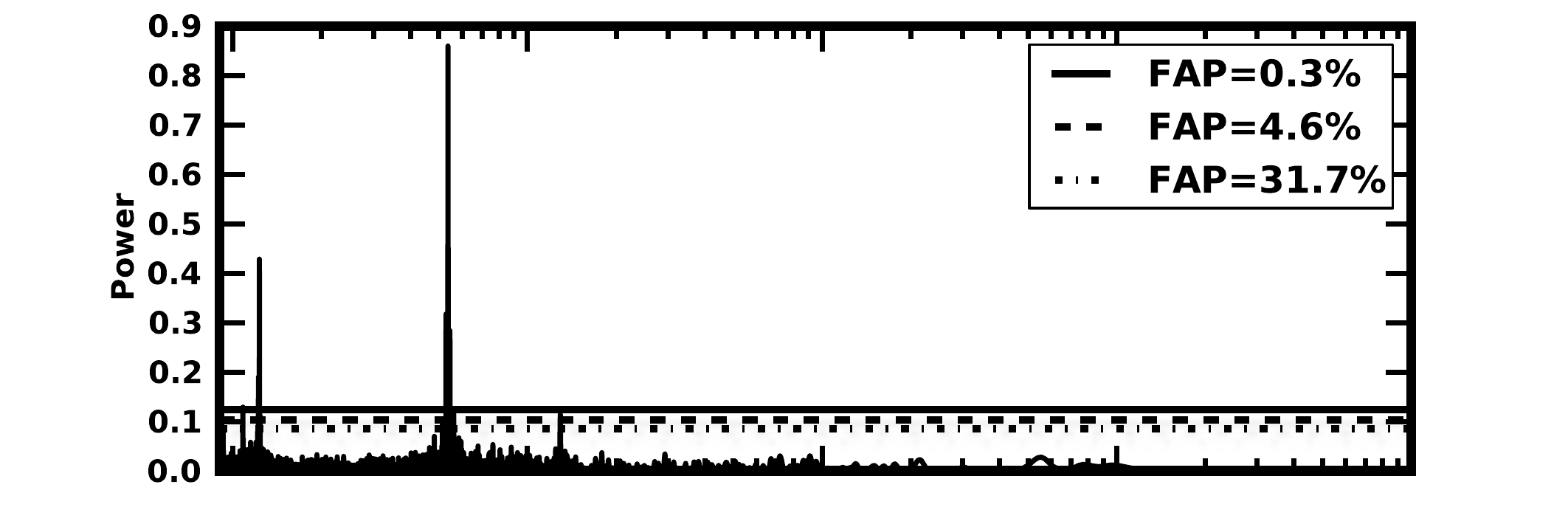}
\includegraphics[width=0.9\linewidth]{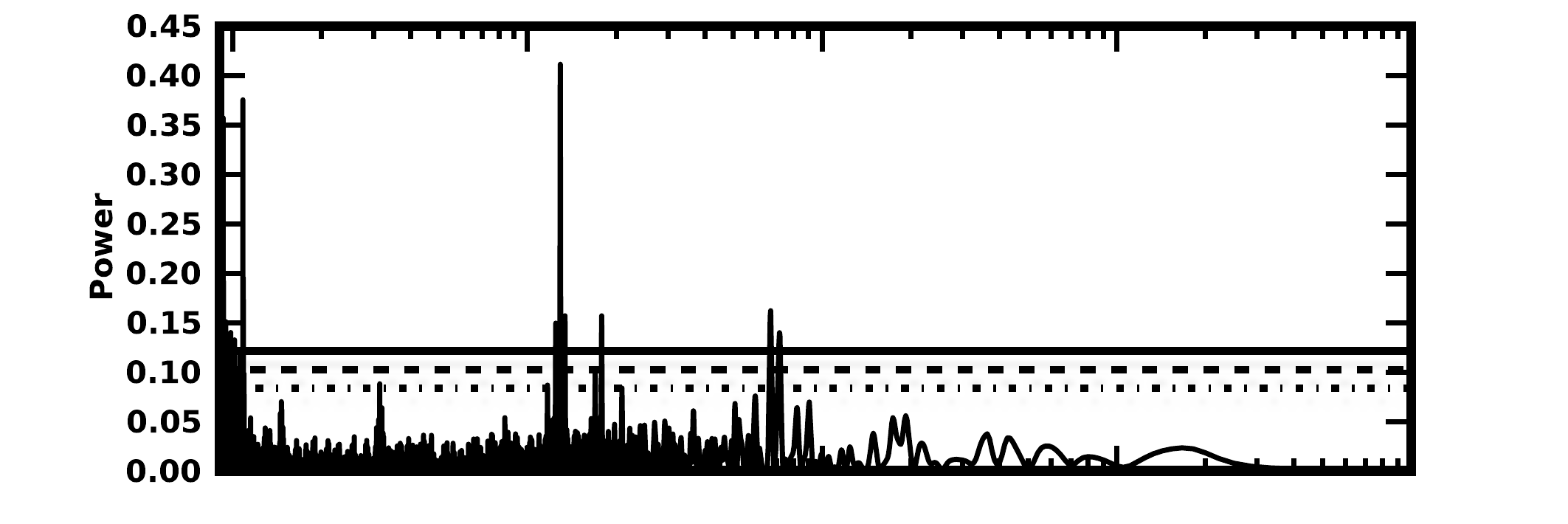}
\includegraphics[width=0.9\linewidth]{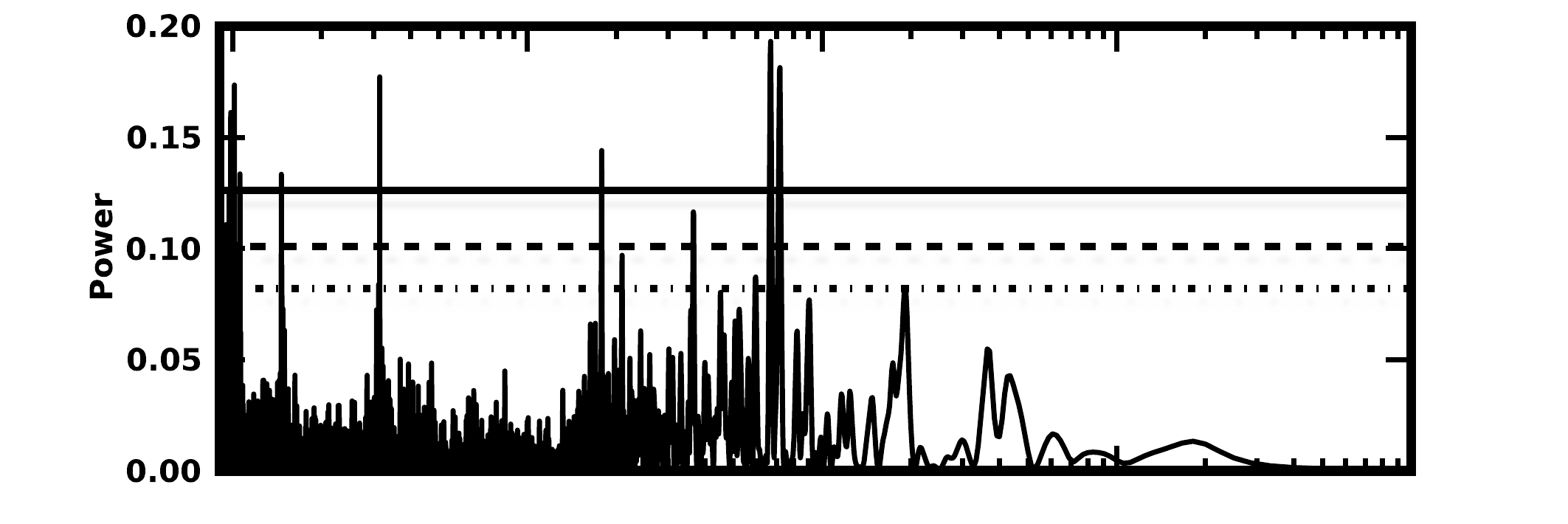}
\includegraphics[width=0.9\linewidth]{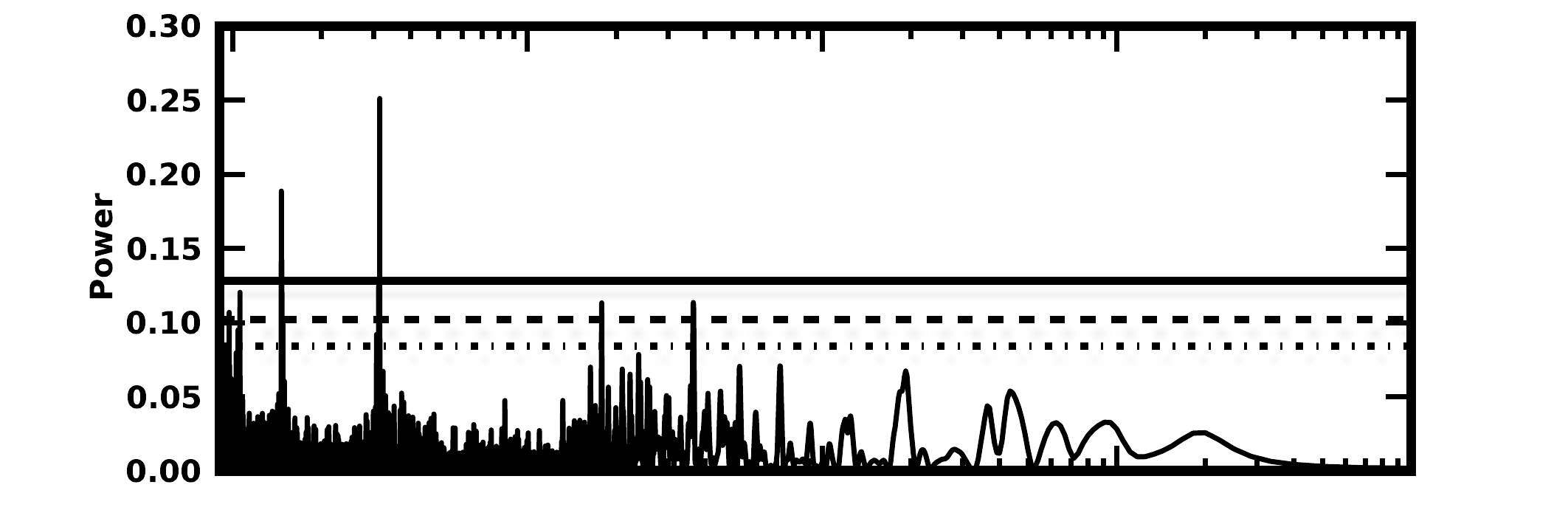}
\includegraphics[width=0.9\linewidth]{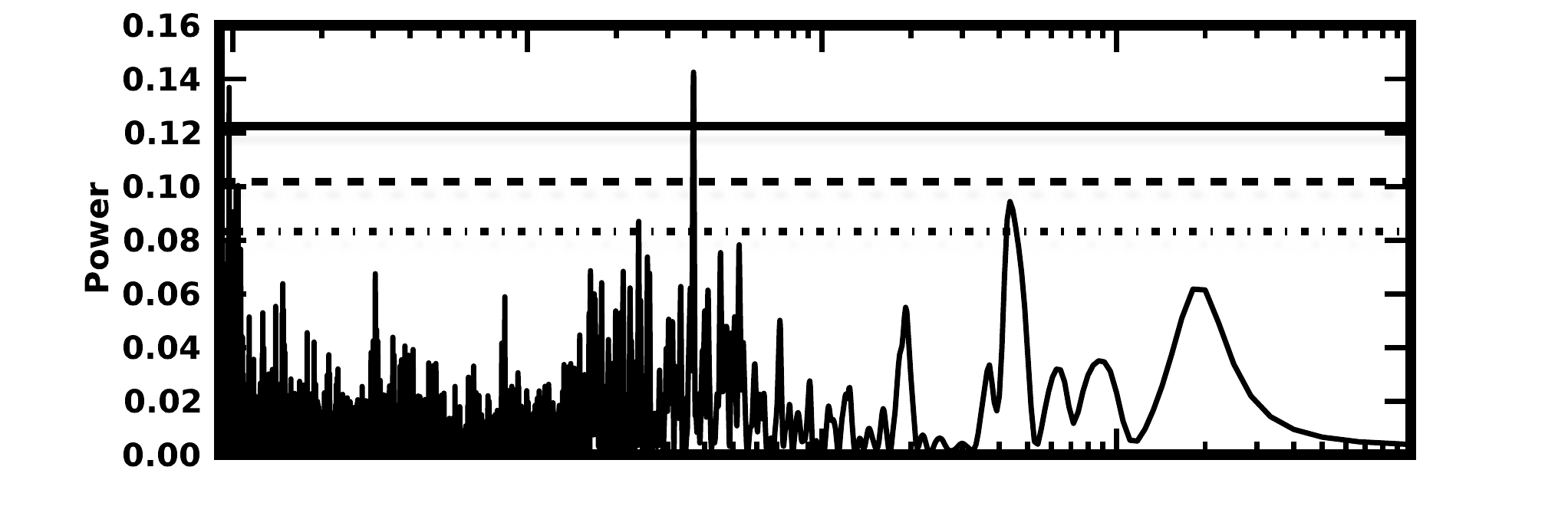}
\includegraphics[width=0.9\linewidth]{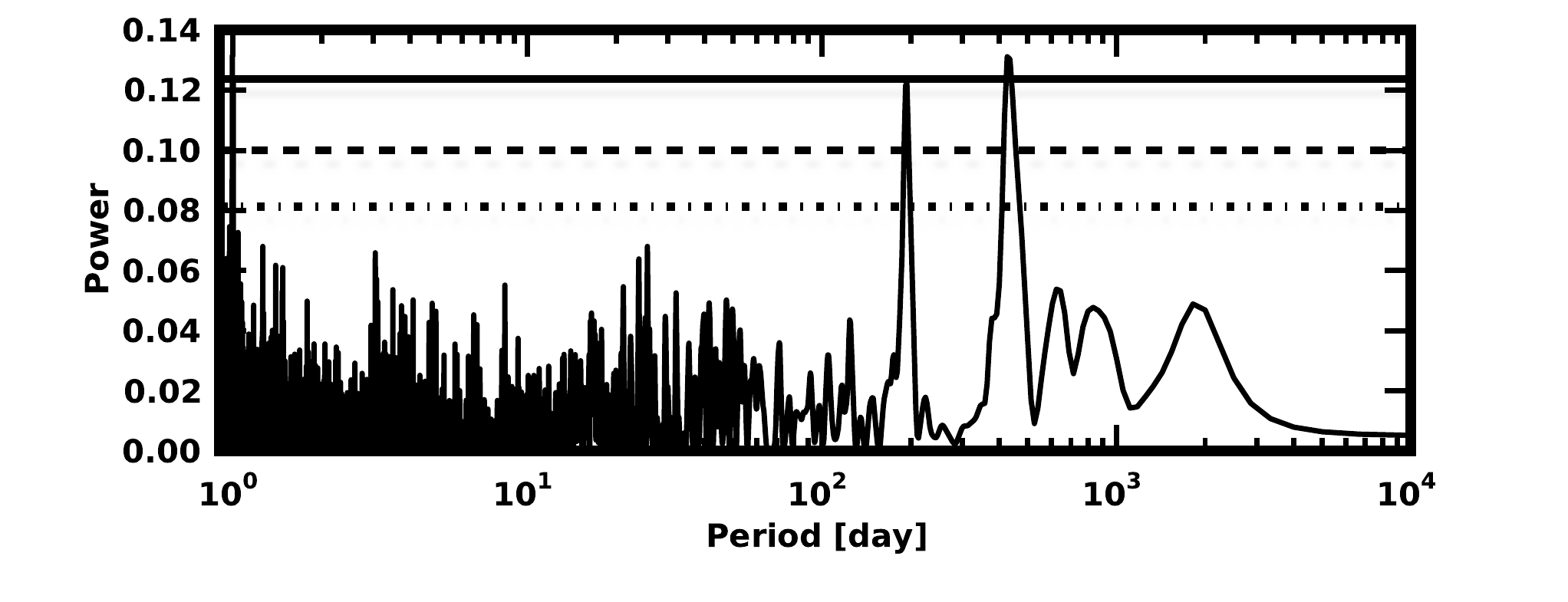}
\caption{
  Equivalent of Fig.~\ref{Fig_Periodogram} for a dataset in which
  the signatures of Gl~581f and Gl~581g (with their \citet{Vogt2010}
  orbital elements) are subtracted from the HARPS measurements. 
  The periodogram of the residuals of the 4-planets orbital model
  has a highly significant peak at the period of Gl~581g and a
  moderately significant peak at the period of Gl~581f, which becomes
  significant after subtracting Gl~581g.
   }
  \label{Fig_PeriodogramVogt}
\end{figure}
To investigate planets with potentially somewhat different 
periods, we turn to a step by step periodogram analysis 
(Fig.~\ref{Fig_Periodogram}), successively removing planets 
with periods initialized at the highest peak of periodogram 
of the previous step. To estimate the False Alarm Probability
(FAP) of those peaks, we use bootstrap resampling 
\citep[e.g.][]{Press1992} of the actual measurements (or residuals)
to generate 10,000 virtual datasets, and adopt the fraction of
those virtual periodograms with a peak above a given power
as the False Alarm Probability for that periodogram power. 
This bootstrap analysis automatically accounts for the actual
level, and the possibly non-Gaussian statistics, of the measurement
noise, but not for any correlation between measurement. It can
therefore potentially underestimate the actual False Alarm levels.
In early iterations the bootstrap analysis on the other hand 
considers the signal from other planets as noise, and therefore 
strongly underestimates significance.
Fig.~\ref{Fig_Periodogram} displays the periodogram power levels 
for False Alarm Probabilities equivalent to 1, 2 and 3~$\sigma$ significance
for a Gaussian statistics. The periodogram analysis, unsurprisingly, 
successively re-identifies Gl~581b to Gl~581e. The highest periodogram 
peak for the last step (Fig.~\ref{Fig_Periodogram}, bottom), which
corresponds to the residuals of the 4-Keplerian model, scarcely 
rises above 1~$\sigma$ significance. The $\sim$50\% excess in 
the residuals is therefore very unlikely to be explained 
by one additional planet. It could reflect either astrophysical 
or instrumental jitter, several planets each contributing 
a signal below our detection threshold, or a combination thereof. 

To evaluate the sensitivity
of our dataset to planets with orbital parameters similar to the
\citet{Vogt2010} Gl~581f and Gl~581g, we repeat the step by step
periodogram analysis for a dataset where we subtract the signatures
of those two planets, with their \citet{Vogt2010} orbital elements,
from the HARPS radial velocities (the choice of subtracting, rather 
than adding, those signatures avoids the suspicion that we could be
boosting a pre-existing signal in the HARPS measurement). The last 
step of that procedure (Fig.~\ref{Fig_PeriodogramVogt}, second panel 
from bottom) 
has a highly significant peak at the period of Gl~581g and a 
moderately significant one at the period of Gl~581f. Removing 
the Gl~581g signal brings the False Alarm Probability of the Gl~581f 
peak down to 0.08\% (Fig.~\ref{Fig_PeriodogramVogt}, bottom panel).
This exercise demonstrates that our dataset is sensitive to such 
planets, and does not find them.

\section{Photometric transit search}
\begin{figure}[h]
\centering
\includegraphics[width=0.9\linewidth]{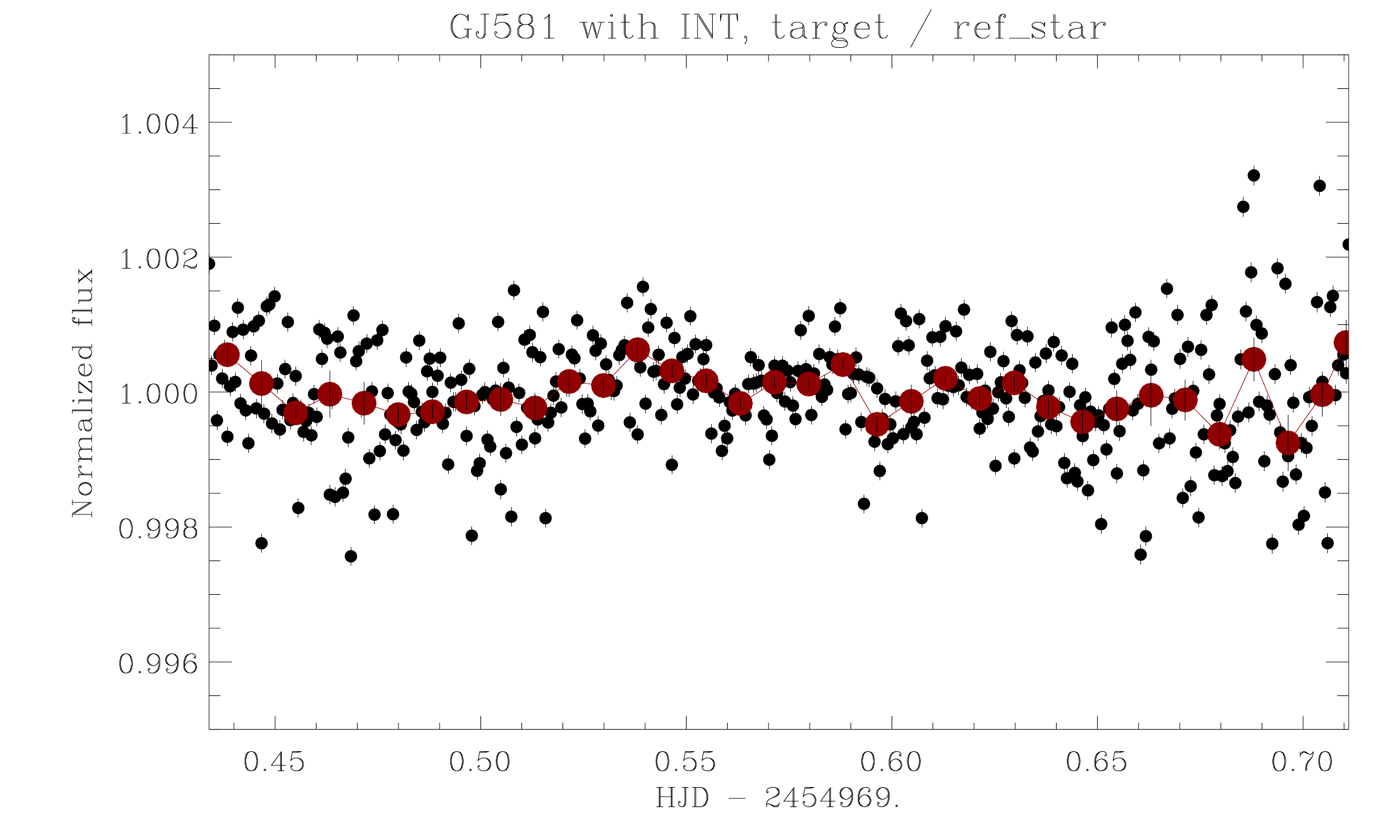}
\caption{r'-band differential light curve of Gl~581 during the potential
  transit of Gl~581e on May~17$^{th}$ 2009, after normalization by
  a third order polynomial. The light curve shows no evidence for 
  a transit. The black points represent the individual measurements, 
  and the red dots their average over 12 min bins. The dispersion 
  of the black points is 830~ppm, and that of the red points is 
  342~ppm. For white noise, the expected dispersion of the 12~mn
  averages would be 230~ppm, revealing 250~ppm of uncorrected 
  systematic red noise on that time-scale.}
\label{fig_lc}
\end{figure}
As mentioned by \citet{Mayor2009}, the {\it a priori} geometric 
probability that Gl~581e transits across its host star is approximately 
{5\%}. Thanks to the small radius of Gl~581, 0.29{\Rsol}, the transit 
of a 1~Earth radius planet across Gl~581 would induce a 1.1~mmag dip 
in the observed flux. Non-grazing transits of the 2-Earth-Mass 
Gl~581e, if they occur at all and unless the planet is much denser 
than the Earth, would thus be detectable with state of the art 
ground-based photometry 
\citep[e.g.][]{Johnson2009b}.

We used the WFC camera \citep{Walton2001} on the 2.5~m Isaac 
Newton Telescope to observe the potential transit on the night of 
May~17$^{th}$ to 18$^{th}$ 2009, with a predicted center at 
JD=2454969.4886$\pm$0.063 (23h43m$\pm$1.52 hrs on 
May~17$^{th}$ 2009). The WFC is a mosaic of 4 2048x4096 EEV CCDs 
which cover a 34'$\times$34' field of view, with an 11'$\times$11' 
gap in one corner. The focal scale is 0.333\arcsec /pixel.
At the time of our observations the detector could neither
be binned nor windowed, and the fastest WFC readout time 
was 29~s. We observed through a Sloan-like r' filter and 
used 25~s exposures, for a 55~s total inter-exposure time. 
To avoid saturating the detector, we defocused the telescope 
to a 15~arcsec image diameter. Since the autoguiding system 
of the WFC does not operate on such severely defocused image, 
the observer guided manually, with measured rms excursions 
of 1.7 and 2.3 pixels (0.6 and 0.8 arcsec) for the X and Y axes. 
We obtained flat field exposures on the twilight
sky at the end of the night, with the telescope defocused by the same 
amount as during the observations, and we used a sequence of dome screen 
exposures with increasing integration times \citep{Gilliland1993} to 
verify the linearity of the CCD detectors. 

\begin{figure}[h]
\centering
\includegraphics[width=0.9\linewidth]{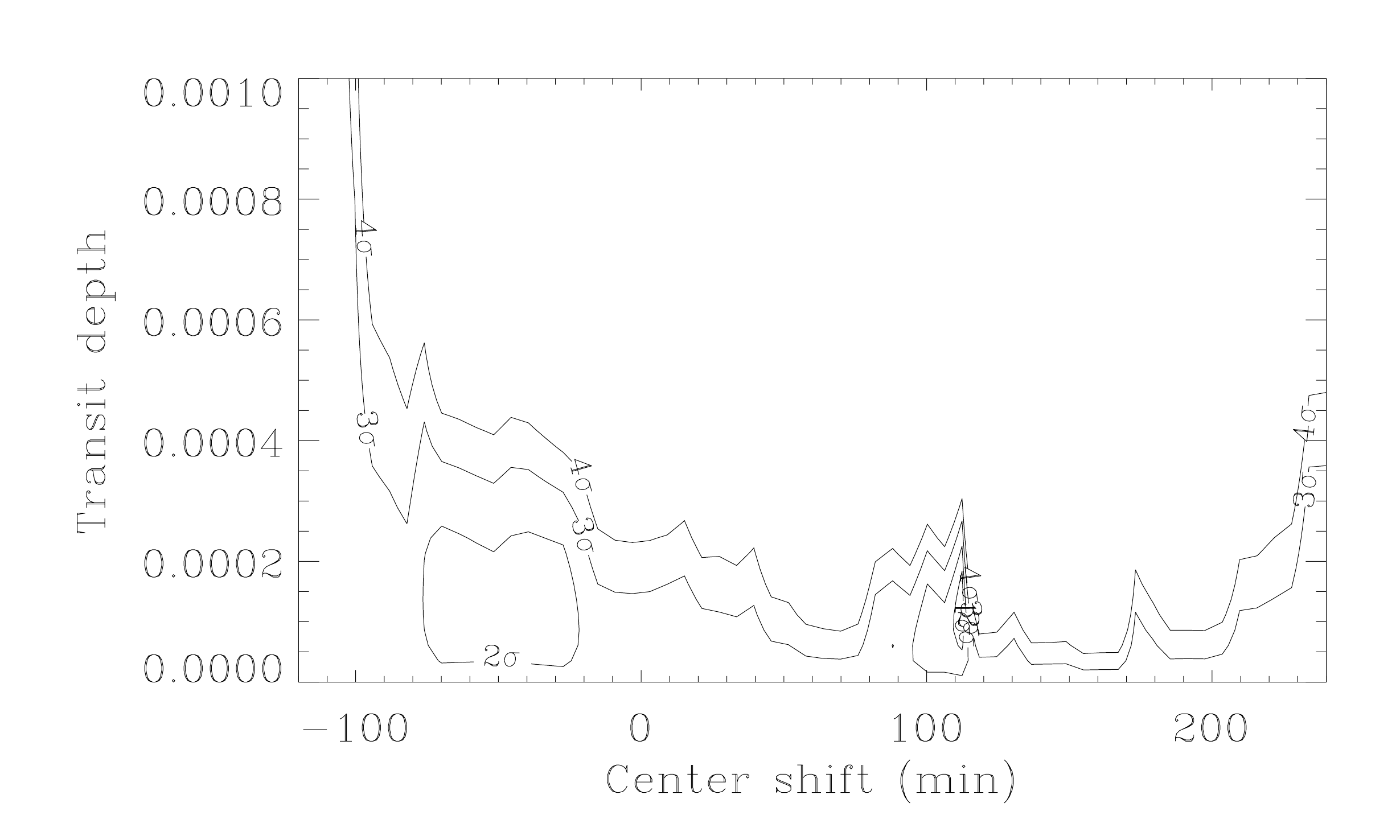}
\caption{Map of the \chisq of the residuals of a fit of a 1~h-long 
  trapeze-shape transit as a function of transit depth and center
  relative to the predicted time (HJD~=~2454969.489).}
\label{fig_map}
\end{figure}

We extracted aperture photometry for the target and several potential
reference stars, using a circular aperture of 50~pixel diameter 
and a sky annulus with 55 and 85~pixels inner and outer diameters. 
The flux of Gl~581 was then divided by the summed fluxes of 4 bright
non variable stars.
The resulting light curve  (Table~\ref{Table_Phot}, only available 
electronically)
shows a low frequency trend, which reflects differential color 
extinction between the (red) target and its reference stars.
That trend is well described by a third order polynomial, 
which is smooth enough to remove signal on the time-scales of
interest for a 1~hour transit only at the edges of the observing
window. After dividing out that trend, the lightcurve 
(Fig.~\ref{fig_lc}) is flat, with a 0.0830\% rms scatter
and no obvious evidence for a transit. 
Finally, we mapped the \chisq residuals of the light-curve with 
respect to a 1~h long trapeze-shape transit as a function of transit 
center and depth. At the 4~$\sigma$ confidence level, the resulting 
map (Figure~\ref{fig_map}) excludes non-grazing transit deeper 
than 0.06\% within -90 to +210~mn of the predicted transit center.
Inserting transits prior to the normalization by the polynomial 
baseline however shows that transits in early part of the transit
window can be partly absorbed into the polynomial baseline, and
that we can exclude 0.06\% deep transit only within -20 to +210~mn 
($-0.22$ to $+2.5~\sigma$) of the predicted center. This depth translates
to a 0.75~\Rearth maximum radius, and therefore to a minimum density 
of 4.4 Earth density, for a 
non-grazing transiting Gl~581e. The planet is therefore most likely
to not transit, but better coverage of the transit window will
be needed to fully exclude that it does. 

\section{Summary}
We observed the Gl~581 planetary system with the HARPS
spectrograph for another two years, adding 121 high-quality 
radial velocity measurements to the 119 which we previously
published \citet{Mayor2009}. The new measurements refine the
parameters of the 4 \citet{Mayor2009} planets, and have in 
particular revised the minimum mass of Gl~581d down to 6~\Mearth. 
The lowered minimum mass of this potentially habitable planet
\citep[e.g.][]{Wordsworth2011} makes a rocky composition more
likely. The revised orbital fit also provides a more precise
ephemeris for recent and future epochs.

When phased to the \citet{Vogt2010} periods for the candidate
Gl~581f and Gl581g planets, the residuals of the 4-planet orbital 
fit show no evidence for a coherent signal with the expected
1.2~m/s amplitude. When we add two planets with elements fixed 
to the \citet{Vogt2010} values for Gl~581f and Gl~581g and readjust
the orbital model, its residuals increase by 75\%. When instead
we fit for two planets with circular orbits of arbitrary amplitude 
and phase at the periods of Gl~581f and  Gl581g in addition to
planets b to e, we obtain amplitudes of 16 and 40~cm/s. In a separate
test, we verified that planets with the characteristics of Gl~581f
and Gl~581g, but for a $\pi$ phase shift, are easily identified
in a step by step periodogram analysis. Our dataset therefore
has strong diagnostic power for planets with the parameters
of Gl~581f and Gl~581g, and we conclude that the Gl~581 system is
unlikely to contain planets with those characteristics. This, of 
course, is not to say that the Gl~581 system can contain no other 
planet, nor even that it contains no planets with periods close to 
33 and 430~days. Both periods correspond to dynamically stable 
orbits \citep[e.g.][]{Vogt2010}, and planets with similar periods
but lower masses or/and different phases certainly could exist
under our noise ceiling. Such planets could potentially contribute 
to the 60\% excess of our residuals over our internal errors,
together with some combination of astrophysical and instrumental
noise. Disentangling those contributions to the excess residuals
may however require significant instrumental or methodological 
improvements, since we already have invested $\sim$10~nights on 
the Gl~581 system and can no longer easily boost our sensitivity
by working up sqrt(N).

We also obtained high-precision (0.08\% rms) relative photometry 
of Gl~581 during a potential transit of the 3-days Gl~581, which
had a $\sim$5\% {\it a priori} geometric transit probability. At
a 4~$\sigma$ confidence level, those observations exclude transits
by a realistically dense Gl~581e for transit centers between -1
and +2.3~$\sigma$ of the date predicted by our ephemeris. Gl~581e 
is thus unlikely to transit, though wider coverage of the transit
window would be needed to fully exclude that it does.

\begin{acknowledgements}

We would like to thank the ESO La Silla staff for their excellent support,
and our collaborators of the HARPS consortium for making this instrument 
such a success, as well as for helping obtain well sampled observations 
through observing time exchange. 

   Some of the reported observations were made with the INT operated on the
island of La Palma by the Isaac Newton Group in the Spanish
Observatorio del Roque de Los Muchachos of the Instituto de 
Astrof{\'\i}sica de Canarias.

NCS acknowledges the support by the European Research 
Council/European Community under the FP7 through Starting Grant 
agreement number 239953, as well as the support from Funda\c{c}\~ao 
para a Ci\^encia e a Tecnologia (FCT) through program Ci\^encia\,2007 
funded by FCT/MCTES (Portugal) and POPH/FSE (EC), and in the form 
of grants reference PTDC/CTE-AST/098528/2008 and PTDC/CTE-AST/098604/2008.

MR acknowledges support from ALMA-CONICYT projects 31090015 and
31080021.

\end{acknowledgements}


\bibliographystyle{aa}
\bibliography{biblio}

\clearpage
\onecolumn
\Online

\begin{table}
\caption{Radial-velocity measurements and error bars for Gl~581
All velocities are relative to the solar system barycenter, and
corrected from the 0.20~[m\,s$^{-1}$] perspective acceleration 
computed from the Hipparcos parallax and proper motion. Only 
available electronically.}
\label{TableRV}
\centering
\begin{longtable}{c c c c}
\hline\hline
{\bf JD-2400000} & {\bf S/N} & {\bf RV} & {\bf Uncertainty} \\
 &  & {\bf [km\,s$^{-1}$]} & {\bf [km\,s$^{-1}$]} \\
\hline
53152.71289 & 48.8 & -9.2163 & 0.0011 \\
53158.66346 & 41.0 & -9.2251 & 0.0013 \\
53511.77334 & 42.8 & -9.2133 & 0.0012 \\
53520.74475 & 38.3 & -9.1957 & 0.0014 \\
53573.51204 & 39.2 & -9.2054 & 0.0013 \\
53574.52233 & 46.0 & -9.1970 & 0.0011 \\
53575.48075 & 55.3 & -9.2017 & 0.0010 \\
53576.53605 & 53.0 & -9.2132 & 0.0010 \\
53577.59260 & 43.0 & -9.2169 & 0.0012 \\
53578.51071 & 62.5 & -9.2057 & 0.0009 \\
53578.62960 & 45.0 & -9.2038 & 0.0011 \\
53579.46256 & 57.8 & -9.1928 & 0.0009 \\
53579.62105 & 49.2 & -9.1913 & 0.0011 \\
53585.46177 & 43.9 & -9.1983 & 0.0011 \\
53586.46516 & 66.2 & -9.2091 & 0.0008 \\
53587.46470 & 30.6 & -9.2233 & 0.0016 \\
53588.53806 & 19.4 & -9.2142 & 0.0026 \\
53589.46202 & 66.7 & -9.2000 & 0.0008 \\
53590.46390 & 59.7 & -9.1933 & 0.0008 \\
53591.46648 & 64.0 & -9.1983 & 0.0008 \\
53592.46481 & 61.5 & -9.2113 & 0.0008 \\
53606.55168 & 24.2 & -9.1912 & 0.0021 \\
53607.50753 & 52.8 & -9.1945 & 0.0010 \\
53608.48264 & 42.7 & -9.2098 & 0.0012 \\
53609.48845 & 32.1 & -9.2174 & 0.0016 \\
53757.87732 & 52.3 & -9.2003 & 0.0010 \\
53760.87548 & 38.9 & -9.2075 & 0.0013 \\
53761.85922 & 37.0 & -9.1986 & 0.0014 \\
53811.84694 & 40.3 & -9.1996 & 0.0013 \\
53813.82702 & 60.7 & -9.2160 & 0.0009 \\
53830.83696 & 56.9 & -9.2078 & 0.0009 \\
53862.70144 & 57.6 & -9.2066 & 0.0009 \\
53864.71366 & 49.1 & -9.1922 & 0.0011 \\
53867.75217 & 48.2 & -9.2153 & 0.0011 \\
53870.69660 & 48.4 & -9.1997 & 0.0011 \\
53882.65776 & 57.9 & -9.2174 & 0.0009 \\
53887.69074 & 66.0 & -9.2117 & 0.0008 \\
53918.62175 & 45.9 & -9.1983 & 0.0011 \\
53920.59495 & 50.7 & -9.2245 & 0.0010 \\
53945.54312 & 51.4 & -9.1968 & 0.0010 \\
53951.48593 & 66.4 & -9.1995 & 0.0008 \\
53975.47160 & 54.9 & -9.2134 & 0.0010 \\
53979.54398 & 41.4 & -9.2134 & 0.0013 \\
54166.87418 & 46.3 & -9.2189 & 0.0011 \\
54170.85396 & 62.1 & -9.1981 & 0.0009 \\
54194.87235 & 48.4 & -9.2185 & 0.0011 \\
54196.75038 & 43.1 & -9.1897 & 0.0012 \\
54197.84504 & 41.9 & -9.1908 & 0.0012 \\
54198.85551 & 40.6 & -9.2082 & 0.0013 \\
54199.73287 & 54.7 & -9.2114 & 0.0010 \\
54200.91092 & 49.5 & -9.2061 & 0.0011 \\
54201.86855 & 52.6 & -9.1967 & 0.0010 \\
54202.88260 & 53.2 & -9.1935 & 0.0010 \\
54228.74156 & 47.9 & -9.1975 & 0.0011 \\
54229.70048 & 33.2 & -9.1957 & 0.0015 \\
54230.76214 & 51.8 & -9.2078 & 0.0010 \\
54234.64592 & 45.1 & -9.1914 & 0.0012 \\
54253.63317 & 50.9 & -9.2154 & 0.0010 \\
54254.66481 & 54.5 & -9.2104 & 0.0010 \\
54291.56885 & 36.7 & -9.2129 & 0.0014 \\
54292.59081 & 62.0 & -9.2058 & 0.0009 \\
54293.62587 & 56.6 & -9.1962 & 0.0010 \\
54295.63945 & 46.3 & -9.2164 & 0.0011 \\
54296.60611 & 40.6 & -9.2257 & 0.0013 \\
54297.64194 & 53.7 & -9.2143 & 0.0010 \\
54298.56760 & 47.8 & -9.1983 & 0.0011 \\
54299.62220 & 31.5 & -9.1951 & 0.0016 \\
54300.61911 & 52.4 & -9.2070 & 0.0010 \\
54315.50749 & 30.5 & -9.1920 & 0.0017 \\
54317.48085 & 56.2 & -9.2130 & 0.0010 \\
54319.49053 & 37.8 & -9.2031 & 0.0014 \\
54320.54407 & 56.9 & -9.1951 & 0.0010 \\
54323.50705 & 14.1 & -9.2171 & 0.0037 \\
54340.55578 & 60.5 & -9.2077 & 0.0009 \\
54342.48620 & 51.1 & -9.1870 & 0.0011 \\
54349.51516 & 52.7 & -9.2179 & 0.0010 \\
54530.85566 & 56.0 & -9.1978 & 0.0010 \\
54550.83127 & 61.7 & -9.1996 & 0.0009 \\
54553.80372 & 72.4 & -9.2189 & 0.0008 \\
54563.83800 & 64.1 & -9.2100 & 0.0009 \\
54566.76115 & 48.9 & -9.2051 & 0.0011 \\
54567.79167 & 57.4 & -9.1967 & 0.0010 \\
54569.79330 & 53.0 & -9.2204 & 0.0010 \\
54570.80425 & 55.7 & -9.2204 & 0.0010 \\
54571.81838 & 49.1 & -9.2019 & 0.0011 \\
54587.86197 & 32.6 & -9.2025 & 0.0016 \\
54588.83880 & 48.8 & -9.1963 & 0.0011 \\
54589.82749 & 49.8 & -9.2000 & 0.0011 \\
54590.81963 & 51.4 & -9.2110 & 0.0010 \\
54591.81712 & 31.6 & -9.2256 & 0.0016 \\
54592.82734 & 60.4 & -9.2158 & 0.0009 \\
54610.74293 & 46.4 & -9.1865 & 0.0011 \\
54611.71348 & 60.7 & -9.1970 & 0.0009 \\
54616.71303 & 37.5 & -9.1973 & 0.0014 \\
54639.68651 & 48.0 & -9.2156 & 0.0011 \\
54640.65441 & 38.9 & -9.2162 & 0.0014 \\
54641.63171 & 53.1 & -9.2044 & 0.0010 \\
54643.64500 & 39.6 & -9.2037 & 0.0013 \\
54644.58703 & 41.0 & -9.2159 & 0.0013 \\
54646.62536 & 43.0 & -9.2131 & 0.0012 \\
54647.57912 & 48.2 & -9.1982 & 0.0011 \\
54648.48482 & 47.7 & -9.1960 & 0.0011 \\
54661.55371 & 44.2 & -9.2177 & 0.0012 \\
54662.54941 & 37.3 & -9.2071 & 0.0014 \\
54663.54487 & 44.7 & -9.1919 & 0.0012 \\
54664.55304 & 41.4 & -9.1928 & 0.0013 \\
54665.56938 & 54.6 & -9.1994 & 0.0010 \\
54672.53172 & 26.8 & -9.2106 & 0.0019 \\
54674.52412 & 38.6 & -9.1967 & 0.0014 \\
54677.50511 & 49.4 & -9.2156 & 0.0011 \\
54678.55679 & 45.0 & -9.2032 & 0.0012 \\
54679.50403 & 31.3 & -9.1928 & 0.0017 \\
54681.51414 & 33.6 & -9.2029 & 0.0015 \\
54682.50334 & 36.9 & -9.2119 & 0.0014 \\
54701.48507 & 42.0 & -9.1919 & 0.0013 \\
54703.51304 & 41.7 & -9.2078 & 0.0013 \\
54708.47905 & 43.6 & -9.2114 & 0.0012 \\
54721.47303 & 40.5 & -9.2205 & 0.0013 \\
54722.47237 & 46.0 & -9.2034 & 0.0012 \\
54916.91735 & 57.0 & -9.2015 & 0.0010 \\
54919.77751 & 59.6 & -9.2200 & 0.0009 \\
54935.69136 & 48.2 & -9.2198 & 0.0011 \\
54938.77023 & 46.7 & -9.2012 & 0.0011 \\
54941.70399 & 54.4 & -9.2185 & 0.0010 \\
54946.74298 & 47.2 & -9.2110 & 0.0011 \\
54955.79358 & 52.1 & -9.2128 & 0.0010 \\
54989.67874 & 55.3 & -9.2116 & 0.0010 \\
54993.61155 & 53.1 & -9.2125 & 0.0010 \\
54998.65589 & 39.5 & -9.2053 & 0.0013 \\
55049.51551 & 42.8 & -9.2063 & 0.0013 \\
55056.52501 & 18.4 & -9.1936 & 0.0029 \\
55227.84095 & 46.4 & -9.1947 & 0.0011 \\
55229.88062 & 36.2 & -9.2118 & 0.0014 \\
55230.85894 & 28.7 & -9.2142 & 0.0018 \\
55232.88302 & 28.6 & -9.1898 & 0.0018 \\
55272.83531 & 46.8 & -9.2089 & 0.0011 \\
55275.80926 & 39.6 & -9.1968 & 0.0013 \\
55277.83041 & 45.4 & -9.2078 & 0.0012 \\
55282.86587 & 44.8 & -9.2046 & 0.0012 \\
55292.84315 & 50.8 & -9.1986 & 0.0011 \\
55294.77402 & 51.7 & -9.2233 & 0.0010 \\
55295.68472 & 40.8 & -9.2195 & 0.0013 \\
55297.76797 & 54.7 & -9.1960 & 0.0010 \\
55298.73452 & 52.1 & -9.2020 & 0.0010 \\
55299.68212 & 30.1 & -9.2132 & 0.0017 \\
55300.72869 & 40.5 & -9.2212 & 0.0013 \\
55301.84323 & 43.4 & -9.2123 & 0.0012 \\
55305.80850 & 54.2 & -9.2187 & 0.0010 \\
55306.76724 & 51.2 & -9.2129 & 0.0010 \\
55307.76067 & 65.9 & -9.1974 & 0.0008 \\
55308.75781 & 57.8 & -9.1883 & 0.0009 \\
55309.76544 & 60.0 & -9.2016 & 0.0009 \\
55321.70852 & 51.6 & -9.2117 & 0.0010 \\
55325.66237 & 39.8 & -9.2006 & 0.0013 \\
55326.61457 & 41.4 & -9.2156 & 0.0013 \\
55328.63743 & 33.4 & -9.2097 & 0.0016 \\
55334.66359 & 39.4 & -9.1920 & 0.0013 \\
55336.78989 & 56.4 & -9.2010 & 0.0010 \\
55337.65473 & 37.5 & -9.2133 & 0.0014 \\
55349.63634 & 17.7 & -9.2022 & 0.0030 \\
55353.57756 & 41.1 & -9.2173 & 0.0013 \\
55354.60681 & 41.7 & -9.2224 & 0.0013 \\
55355.53696 & 42.4 & -9.2072 & 0.0013 \\
55359.56247 & 44.2 & -9.2138 & 0.0012 \\
55370.57818 & 26.3 & -9.2175 & 0.0020 \\
55372.55366 & 34.3 & -9.1934 & 0.0015 \\
55373.60234 & 46.0 & -9.1943 & 0.0012 \\
55374.61617 & 35.2 & -9.2084 & 0.0015 \\
55375.55663 & 39.3 & -9.2155 & 0.0014 \\
55389.64756 & 33.9 & -9.1941 & 0.0015 \\
55390.54432 & 11.0 & -9.2014 & 0.0047 \\
55391.54670 & 39.1 & -9.2193 & 0.0014 \\
55396.49708 & 25.3 & -9.2145 & 0.0021 \\
55399.54017 & 42.2 & -9.1874 & 0.0013 \\
55401.52230 & 30.9 & -9.2063 & 0.0017 \\
55407.49699 & 54.2 & -9.2174 & 0.0010 \\
55408.50168 & 27.2 & -9.2111 & 0.0019 \\
55410.55603 & 43.5 & -9.1892 & 0.0012 \\
55411.51484 & 34.7 & -9.1973 & 0.0015 \\
55423.51171 & 48.7 & -9.2171 & 0.0011 \\
55427.49846 & 48.1 & -9.1975 & 0.0011 \\
55428.48093 & 40.5 & -9.2108 & 0.0013 \\
55434.51127 & 46.3 & -9.2217 & 0.0012 \\
55435.48705 & 56.7 & -9.2177 & 0.0010 \\
55436.48340 & 60.2 & -9.2024 & 0.0009 \\
55437.51432 & 59.7 & -9.1912 & 0.0009 \\
55439.48708 & 53.5 & -9.2156 & 0.0010 \\
55443.49986 & 24.8 & -9.2031 & 0.0021 \\
55444.48950 & 47.8 & -9.2166 & 0.0011 \\
55445.49328 & 44.9 & -9.2280 & 0.0012 \\
55450.48002 & 39.1 & -9.2164 & 0.0014 \\
55453.48660 & 46.7 & -9.1912 & 0.0012 \\
55454.47680 & 38.0 & -9.2019 & 0.0014 \\
55455.48896 & 40.2 & -9.2155 & 0.0013 \\
55457.47397 & 52.6 & -9.2104 & 0.0010 \\
55458.48996 & 40.9 & -9.1994 & 0.0013 \\
55464.48161 & 31.2 & -9.1896 & 0.0017 \\
55626.90847 & 31.2 & -9.2099 & 0.0017 \\
55627.86994 & 44.2 & -9.2227 & 0.0012 \\
55629.88250 & 54.3 & -9.1989 & 0.0010 \\
55630.88945 & 46.2 & -9.1915 & 0.0012 \\
55633.83855 & 58.9 & -9.2171 & 0.0009 \\
55634.83780 & 48.9 & -9.2046 & 0.0011 \\
55635.80037 & 39.5 & -9.1911 & 0.0014 \\
55638.87580 & 49.5 & -9.2200 & 0.0011 \\
55639.82564 & 56.6 & -9.2145 & 0.0010 \\
55641.85816 & 54.3 & -9.1963 & 0.0010 \\
55642.78865 & 43.5 & -9.2068 & 0.0012 \\
55644.87268 & 50.3 & -9.2115 & 0.0011 \\
55646.85119 & 43.2 & -9.1931 & 0.0012 \\
55647.86060 & 55.1 & -9.2042 & 0.0010 \\
55648.89760 & 50.0 & -9.2160 & 0.0011 \\
55652.83978 & 51.6 & -9.2023 & 0.0011 \\
55653.72224 & 50.5 & -9.2147 & 0.0011 \\
55654.68243 & 50.3 & -9.2219 & 0.0011 \\
55656.75878 & 57.5 & -9.1999 & 0.0010 \\
55657.76630 & 51.4 & -9.1918 & 0.0011 \\
55658.82034 & 34.5 & -9.2014 & 0.0015 \\
55662.76574 & 36.8 & -9.1915 & 0.0014 \\
55663.75875 & 17.3 & -9.1988 & 0.0030 \\
55672.71848 & 46.1 & -9.1978 & 0.0012 \\
55674.73187 & 52.8 & -9.2027 & 0.0010 \\
55675.77838 & 40.9 & -9.2182 & 0.0013 \\
55676.75948 & 39.0 & -9.2151 & 0.0014 \\
55677.69188 & 34.2 & -9.2023 & 0.0015 \\
55678.76651 & 35.5 & -9.1948 & 0.0015 \\
55679.71364 & 58.7 & -9.1996 & 0.0009 \\
55680.62402 & 41.9 & -9.2082 & 0.0013 \\
55681.67631 & 54.6 & -9.2190 & 0.0010 \\
55682.67267 & 52.5 & -9.2083 & 0.0010 \\
55683.62142 & 54.1 & -9.1926 & 0.0010 \\
55684.67393 & 50.7 & -9.1917 & 0.0011 \\
55685.64645 & 62.9 & -9.2045 & 0.0009 \\
55686.65353 & 36.9 & -9.2133 & 0.0014 \\
55689.71145 & 40.6 & -9.1935 & 0.0013 \\
55690.73996 & 33.8 & -9.2072 & 0.0016 \\
55691.69250 & 36.3 & -9.2180 & 0.0014 \\
55692.71193 & 46.3 & -9.2188 & 0.0012 \\
55693.75657 & 54.6 & -9.2083 & 0.0010 \\
55695.62767 & 58.7 & -9.1939 & 0.0009 \\

\hline
\end{longtable}
\end{table}

\begin{table}
\caption{Relative photometry of Gl~581 during the 
potential May 17$^{th}$ transit of Gl~581e, before 
correction of the slow drits by a polynomial.
Only available electronically.}
\label{Table_Phot}
\centering
\begin{tabular}{c c c}
\hline\hline
{\bf JD-2400000} & {\bf r'} \\
\hline
54969.43393   &   1.003231  \\ 
54969.43457   &   1.001711  \\ 
54969.43522   &   1.002289  \\ 
54969.43585   &   1.000873  \\ 
54969.43649   &   1.001837  \\ 
54969.43713   &   1.001473  \\ 
54969.43776   &   1.001896  \\ 
54969.43841   &   1.000586  \\ 
54969.43905   &   1.001324  \\ 
54969.43969   &   1.002118  \\ 
54969.44032   &   1.001365  \\ 
54969.44095   &   1.002456  \\ 
54969.44160   &   1.001024  \\ 
54969.44223   &   1.002106  \\ 
54969.44288   &   1.000896  \\ 
54969.44352   &   1.000395  \\ 
54969.44415   &   1.001685  \\ 
54969.44479   &   1.002107  \\ 
54969.44543   &   1.000877  \\ 
54969.44606   &   1.002166  \\ 
54969.44671   &   0.998851  \\ 
54969.44735   &   1.000762  \\ 
54969.44798   &   1.002343  \\ 
54969.44862   &   1.002358  \\ 
54969.44925   &   1.000580  \\ 
54969.44989   &   1.002454  \\ 
54969.45054   &   1.001147  \\ 
54969.45118   &   1.000453  \\ 
54969.45183   &   1.000730  \\ 
54969.45246   &   1.001324  \\ 
54969.45310   &   1.002008  \\ 
54969.45372   &   1.000541  \\ 
54969.45437   &   1.000782  \\ 
54969.45501   &   1.001169  \\ 
54969.45565   &   0.999198  \\ 
54969.45629   &   1.000649  \\ 
54969.45692   &   1.000302  \\ 
54969.45757   &   1.000450  \\ 
54969.45820   &   1.000557  \\ 
54969.45884   &   1.000211  \\ 
54969.45949   &   1.000808  \\ 
54969.46013   &   1.000459  \\ 
54969.46076   &   1.001742  \\ 
54969.46140   &   1.001292  \\ 
54969.46205   &   1.001665  \\ 
54969.46268   &   1.001565  \\ 
54969.46331   &   0.999238  \\ 
54969.46396   &   1.000638  \\ 
54969.46460   &   0.999178  \\ 
54969.46524   &   1.001544  \\ 
54969.46588   &   1.001292  \\ 
54969.46651   &   0.999200  \\ 
54969.46715   &   0.999394  \\ 
54969.46780   &   0.999992  \\ 
54969.46845   &   0.998215  \\ 
54969.46908   &   1.001774  \\ 
54969.46972   &   1.001082  \\ 
54969.47036   &   1.001217  \\ 
54969.47099   &   1.000355  \\ 
54969.47164   &   1.000418  \\ 
54969.47228   &   1.001288  \\ 
54969.47291   &   0.999573  \\ 
54969.47355   &   1.000552  \\ 
54969.47419   &   0.998711  \\ 
54969.47484   &   1.001283  \\ 
54969.47547   &   0.999626  \\ 
54969.47612   &   1.001408  \\ 
54969.47675   &   0.999846  \\ 
54969.47739   &   1.000445  \\ 
54969.47803   &   1.000111  \\ 
54969.47868   &   0.998621  \\ 
54969.47932   &   0.999706  \\ 
54969.47995   &   0.999958  \\ 
54969.48058   &   0.999913  \\ 
54969.48122   &   0.999509  \\ 
54969.48187   &   1.000878  \\ 
54969.48251   &   1.000357  \\ 
54969.48314   &   1.000311  \\ 
54969.48378   &   1.000175  \\ 
54969.48443   &   0.999748  \\ 
54969.48507   &   1.001058  \\ 
54969.48570   &   0.999991  \\ 
54969.48633   &   0.999822  \\ 
54969.48698   &   1.000563  \\ 
54969.48761   &   1.000737  \\ 
54969.48826   &   1.000230  \\ 
54969.48890   &   0.999916  \\ 
54969.48953   &   1.000443  \\ 
54969.49017   &   1.000696  \\ 
54969.49081   &   0.999704  \\ 
54969.49146   &   0.999852  \\ 
54969.49209   &   0.999073  \\ 
54969.49274   &   1.000273  \\ 
54969.49338   &   0.999974  \\ 
54969.49401   &   0.999961  \\ 
54969.49465   &   1.001111  \\ 
54969.49529   &   1.000262  \\ 
54969.49594   &   0.999890  \\ 
54969.49657   &   0.999402  \\ 
54969.49721   &   1.000385  \\ 
54969.49785   &   0.997897  \\ 
54969.49849   &   0.999807  \\ 
54969.49913   &   0.998833  \\ 
54969.49977   &   0.998933  \\ 
54969.50040   &   0.999936  \\ 
54969.50105   &   0.999960  \\ 
54969.50168   &   0.999243  \\ 
54969.50232   &   0.999127  \\ 
54969.50297   &   0.999814  \\ 
54969.50361   &   0.999932  \\ 
54969.50425   &   1.000937  \\ 
54969.50489   &   0.998442  \\ 
54969.50552   &   1.000230  \\ 
54969.50617   &   0.998954  \\ 
54969.50680   &   0.999913  \\ 
54969.50745   &   0.997987  \\ 
54969.50810   &   1.001330  \\ 
54969.50873   &   0.999294  \\ 
54969.50937   &   0.999784  \\ 
54969.51002   &   0.999005  \\ 
54969.51066   &   1.000549  \\ 
54969.51129   &   0.999525  \\ 
54969.51194   &   1.000594  \\ 
54969.51258   &   1.000327  \\ 
54969.51323   &   0.999039  \\ 
54969.51386   &   0.999307  \\ 
54969.51450   &   1.000218  \\ 
54969.51515   &   1.000877  \\ 
54969.51578   &   0.997809  \\ 
54969.51642   &   0.999214  \\ 
54969.51707   &   0.999348  \\ 
54969.51771   &   0.999590  \\ 
54969.51834   &   0.999788  \\ 
54969.51898   &   1.000256  \\ 
54969.51962   &   0.999381  \\ 
54969.52028   &   0.999776  \\ 
54969.52091   &   0.999657  \\ 
54969.52155   &   0.999643  \\ 
54969.52218   &   1.000120  \\ 
54969.52282   &   1.000054  \\ 
54969.52346   &   1.000607  \\ 
54969.52409   &   0.999734  \\ 
54969.52474   &   0.999344  \\ 
54969.52539   &   0.998821  \\ 
54969.52602   &   0.999331  \\ 
54969.52666   &   0.999206  \\ 
54969.52731   &   1.000322  \\ 
54969.52795   &   1.000083  \\ 
54969.52859   &   0.998964  \\ 
54969.52922   &   1.000168  \\ 
54969.52986   &   0.999488  \\ 
54969.53051   &   0.999842  \\ 
54969.53114   &   0.999602  \\ 
54969.53179   &   0.999438  \\ 
54969.53243   &   0.999420  \\ 
54969.53306   &   0.999491  \\ 
54969.53371   &   0.999932  \\ 
54969.53435   &   1.000003  \\ 
54969.53499   &   1.000060  \\ 
54969.53562   &   1.000680  \\ 
54969.53627   &   0.998900  \\ 
54969.53691   &   0.999982  \\ 
54969.53754   &   0.999872  \\ 
54969.53818   &   0.998695  \\ 
54969.53883   &   0.999680  \\ 
54969.53946   &   1.000866  \\ 
54969.54010   &   0.999127  \\ 
54969.54074   &   1.000247  \\ 
54969.54137   &   1.000512  \\ 
54969.54202   &   0.999577  \\ 
54969.54266   &   0.999582  \\ 
54969.54330   &   0.999813  \\ 
54969.54393   &   0.998913  \\ 
54969.54458   &   0.999363  \\ 
54969.54522   &   1.000265  \\ 
54969.54585   &   0.999588  \\ 
54969.54649   &   0.998147  \\ 
54969.54714   &   1.000019  \\ 
54969.54777   &   0.999029  \\ 
54969.54842   &   0.999270  \\ 
54969.54905   &   0.999718  \\ 
54969.54970   &   0.999392  \\ 
54969.55034   &   0.999752  \\ 
54969.55098   &   1.000304  \\ 
54969.55163   &   0.999139  \\ 
54969.55227   &   0.999878  \\ 
54969.55291   &   0.999330  \\ 
54969.55354   &   0.999368  \\ 
54969.55419   &   0.999640  \\ 
54969.55483   &   0.999843  \\ 
54969.55547   &   0.999330  \\ 
54969.55611   &   0.998523  \\ 
54969.55676   &   0.999124  \\ 
54969.55739   &   0.999182  \\ 
54969.55803   &   0.999046  \\ 
54969.55867   &   0.998249  \\ 
54969.55931   &   0.998617  \\ 
54969.55995   &   0.998426  \\ 
54969.56059   &   0.998959  \\ 
54969.56122   &   0.998832  \\ 
54969.56187   &   0.999078  \\ 
54969.56251   &   0.998926  \\ 
54969.56315   &   0.998928  \\ 
54969.56378   &   0.998823  \\ 
54969.56443   &   0.998745  \\ 
54969.56507   &   0.999206  \\ 
54969.56570   &   0.999602  \\ 
54969.56635   &   0.999209  \\ 
54969.56699   &   0.999219  \\ 
54969.56762   &   0.999478  \\ 
54969.56826   &   0.999246  \\ 
54969.56889   &   0.998729  \\ 
54969.56953   &   0.998669  \\ 
54969.57018   &   0.998071  \\ 
54969.57082   &   0.998425  \\ 
54969.57146   &   0.999463  \\ 
54969.57209   &   0.999057  \\ 
54969.57273   &   0.999352  \\ 
54969.57338   &   0.999460  \\ 
54969.57401   &   0.998930  \\ 
54969.57465   &   0.999386  \\ 
54969.57529   &   0.999216  \\ 
54969.57593   &   0.998870  \\ 
54969.57657   &   0.998670  \\ 
54969.57721   &   0.999385  \\ 
54969.57784   &   0.999988  \\ 
54969.57849   &   0.999192  \\ 
54969.57913   &   0.999414  \\ 
54969.57978   &   1.000204  \\ 
54969.58041   &   0.998739  \\ 
54969.58105   &   0.999355  \\ 
54969.58169   &   0.999128  \\ 
54969.58233   &   0.999648  \\ 
54969.58297   &   0.999016  \\ 
54969.58362   &   0.999201  \\ 
54969.58426   &   0.999126  \\ 
54969.58489   &   0.999614  \\ 
54969.58553   &   0.999506  \\ 
54969.58617   &   1.000073  \\ 
54969.58681   &   0.999593  \\ 
54969.58746   &   1.000352  \\ 
54969.58810   &   0.999553  \\ 
54969.58874   &   0.998567  \\ 
54969.58937   &   0.999094  \\ 
54969.59002   &   0.999113  \\ 
54969.59066   &   0.999402  \\ 
54969.59130   &   0.999651  \\ 
54969.59193   &   0.999404  \\ 
54969.59258   &   0.998705  \\ 
54969.59321   &   0.997500  \\ 
54969.59385   &   0.999389  \\ 
54969.59450   &   0.999381  \\ 
54969.59514   &   0.998639  \\ 
54969.59578   &   0.998446  \\ 
54969.59642   &   0.999242  \\ 
54969.59706   &   0.998026  \\ 
54969.59769   &   0.998809  \\ 
54969.59834   &   0.999108  \\ 
54969.59899   &   0.998447  \\ 
54969.59963   &   0.998748  \\ 
54969.60026   &   0.998549  \\ 
54969.60091   &   0.999106  \\ 
54969.60155   &   0.999932  \\ 
54969.60218   &   1.000424  \\ 
54969.60281   &   0.998646  \\ 
54969.60346   &   1.000330  \\ 
54969.60409   &   0.999981  \\ 
54969.60473   &   0.998777  \\ 
54969.60538   &   0.998867  \\ 
54969.60601   &   0.998695  \\ 
54969.60665   &   1.000411  \\ 
54969.60730   &   0.997475  \\ 
54969.60794   &   0.998972  \\ 
54969.60857   &   0.999828  \\ 
54969.60921   &   0.999579  \\ 
54969.60986   &   1.000199  \\ 
54969.61049   &   0.999649  \\ 
54969.61112   &   0.999321  \\ 
54969.61178   &   1.000244  \\ 
54969.61242   &   0.999330  \\ 
54969.61305   &   1.000427  \\ 
54969.61369   &   0.999559  \\ 
54969.61434   &   0.999657  \\ 
54969.61497   &   0.999567  \\ 
54969.61561   &   1.000406  \\ 
54969.61626   &   0.999622  \\ 
54969.61689   &   0.999898  \\ 
54969.61754   &   1.000773  \\ 
54969.61818   &   0.999493  \\ 
54969.61882   &   0.999838  \\ 
54969.61945   &   0.999484  \\ 
54969.62009   &   0.999517  \\ 
54969.62074   &   0.999899  \\ 
54969.62137   &   0.999105  \\ 
54969.62202   &   0.999359  \\ 
54969.62266   &   0.999702  \\ 
54969.62329   &   0.999299  \\ 
54969.62394   &   1.000312  \\ 
54969.62458   &   0.999488  \\ 
54969.62522   &   0.998658  \\ 
54969.62586   &   0.999768  \\ 
54969.62650   &   0.999935  \\ 
54969.62715   &   1.000268  \\ 
54969.62779   &   0.999736  \\ 
54969.62843   &   0.999481  \\ 
54969.62913   &   1.000923  \\ 
54969.62978   &   0.998906  \\ 
54969.63041   &   1.000757  \\ 
54969.63106   &   1.000255  \\ 
54969.63170   &   1.000089  \\ 
54969.63233   &   0.999889  \\ 
54969.63298   &   1.000823  \\ 
54969.63361   &   0.999196  \\ 
54969.63426   &   0.999160  \\ 
54969.63489   &   1.000502  \\ 
54969.63553   &   0.999503  \\ 
54969.63617   &   0.999753  \\ 
54969.63681   &   1.000000  \\ 
54969.63745   &   1.000728  \\ 
54969.63809   &   1.000206  \\ 
54969.63873   &   0.999726  \\ 
54969.63937   &   1.000973  \\ 
54969.64002   &   0.999749  \\ 
54969.64066   &   1.000059  \\ 
54969.64129   &   1.000848  \\ 
54969.64193   &   0.999281  \\ 
54969.64258   &   0.999080  \\ 
54969.64321   &   1.000389  \\ 
54969.64386   &   1.000831  \\ 
54969.64450   &   0.999241  \\ 
54969.64513   &   0.999142  \\ 
54969.64577   &   1.000299  \\ 
54969.64641   &   0.999870  \\ 
54969.64704   &   1.000474  \\ 
54969.64769   &   0.999122  \\ 
54969.64834   &   1.000320  \\ 
54969.64898   &   0.999630  \\ 
54969.64961   &   1.000227  \\ 
54969.65026   &   1.000353  \\ 
54969.65090   &   0.998770  \\ 
54969.65154   &   1.000270  \\ 
54969.65218   &   0.999941  \\ 
54969.65282   &   1.000574  \\ 
54969.65346   &   1.001812  \\ 
54969.65409   &   1.001082  \\ 
54969.65473   &   0.999711  \\ 
54969.65538   &   1.000371  \\ 
54969.65601   &   1.001405  \\ 
54969.65667   &   1.002018  \\ 
54969.65731   &   1.001813  \\ 
54969.65796   &   1.001565  \\ 
54969.65860   &   1.000850  \\ 
54969.65923   &   1.002339  \\ 
54969.65987   &   1.001019  \\ 
54969.66050   &   0.998813  \\ 
54969.66114   &   1.000100  \\ 
54969.66178   &   0.999159  \\ 
54969.66243   &   1.002155  \\ 
54969.66306   &   1.001705  \\ 
54969.66370   &   1.002163  \\ 
54969.66435   &   1.001401  \\ 
54969.66498   &   1.000725  \\ 
54969.66562   &   0.991208  \\ 
54969.66627   &   1.018553  \\ 
54969.66691   &   1.003136  \\ 
54969.66755   &   1.000953  \\ 
54969.66819   &   1.001434  \\ 
54969.66882   &   1.001626  \\ 
54969.66946   &   1.002906  \\ 
54969.67009   &   1.002294  \\ 
54969.67074   &   1.000270  \\ 
54969.67137   &   1.001768  \\ 
54969.67201   &   1.002598  \\ 
54969.67266   &   1.000570  \\ 
54969.67329   &   1.002029  \\ 
54969.67393   &   1.001157  \\ 
54969.67458   &   1.000236  \\ 
54969.67522   &   1.002768  \\ 
54969.67586   &   1.001558  \\ 
54969.67650   &   1.003375  \\ 
54969.67714   &   1.002537  \\ 
54969.67777   &   1.003613  \\ 
54969.67842   &   1.001131  \\ 
54969.67906   &   1.002065  \\ 
54969.67969   &   1.002281  \\ 
54969.68033   &   1.001256  \\ 
54969.68098   &   1.001793  \\ 
54969.68163   &   1.001426  \\ 
54969.68226   &   1.002081  \\ 
54969.68290   &   1.001732  \\ 
54969.68354   &   1.001393  \\ 
54969.68419   &   1.002429  \\ 
54969.68483   &   1.003332  \\ 
54969.68547   &   1.005649  \\ 
54969.68611   &   1.004144  \\ 
54969.68674   &   1.002692  \\ 
54969.68738   &   1.004826  \\ 
54969.68803   &   1.006320  \\ 
54969.68866   &   1.004148  \\ 
54969.68931   &   1.003059  \\ 
54969.68995   &   1.004127  \\ 
54969.69058   &   1.002279  \\ 
54969.69122   &   1.003161  \\ 
54969.69187   &   1.003080  \\ 
54969.69251   &   1.001215  \\ 
54969.69314   &   1.003791  \\ 
54969.69379   &   1.005422  \\ 
54969.69443   &   1.003035  \\ 
54969.69506   &   1.002359  \\ 
54969.69571   &   1.005360  \\ 
54969.69636   &   1.002851  \\ 
54969.69699   &   1.004262  \\ 
54969.69763   &   1.003761  \\ 
54969.69827   &   1.002751  \\ 
54969.69892   &   1.002063  \\ 
54969.69956   &   1.003339  \\ 
54969.70019   &   1.002312  \\ 
54969.70084   &   1.003384  \\ 
54969.70147   &   1.004198  \\ 
54969.70211   &   1.003832  \\ 
54969.70276   &   1.004359  \\ 
54969.70340   &   1.005798  \\ 
54969.70405   &   1.007592  \\ 
54969.70469   &   1.004737  \\ 
54969.70533   &   1.003152  \\ 
54969.70598   &   1.002460  \\ 
54969.70662   &   1.006039  \\ 
54969.70725   &   1.006271  \\ 
54969.70789   &   1.004855  \\ 
54969.70854   &   1.005365  \\ 
54969.70917   &   1.005771  \\ 
54969.70981   &   1.005656  \\ 
54969.71046   &   1.005445  \\ 
54969.71109   &   1.007430  \\ 

\hline
\end{tabular}
\end{table}

\end{document}